\documentclass[useAMS,usenatbib]{aastex61}

\submitjournal{ApJ}
\usepackage{enumitem}
\usepackage{calc}
\usepackage{chngcntr}
\usepackage{epsfig}
\usepackage{amsmath}
\usepackage{natbib}
\usepackage{graphicx}
\usepackage[caption=false]{subfig}
\usepackage{wrapfig}
\usepackage[T1]{fontenc}
\usepackage{enumitem}
\usepackage[normalem]{ulem}
\usepackage{textcomp}
\usepackage{longtable}
\usepackage{answers}
\usepackage{array}
\usepackage{tablefootnote}
\usepackage{float}
\newcommand{\beq}{\begin{equation}}
\newcommand{\eeq}{\end{equation}}

\begin{document}

\shorttitle{JCMT Transient Data Reduction/Calibration}
\shortauthors{Mairs et al.}
\title{The JCMT Transient Survey: Data Reduction and Calibration Methods}

\correspondingauthor{Steve Mairs}
\email{smairs@uvic.ca}

\author{Steve Mairs}
\affiliation{Department of Physics and Astronomy, University of Victoria, Victoria, BC, V8P 1A1, Canada}
\affiliation{NRC Herzberg Astronomy and Astrophysics, 5071 West Saanich Rd, Victoria, BC, V9E 2E7, Canada}

\author{James Lane}
\affiliation{Department of Physics and Astronomy, University of Victoria, Victoria, BC, V8P 1A1, Canada}

\author{Doug Johnstone}
\affiliation{NRC Herzberg Astronomy and Astrophysics, 5071 West Saanich Rd, Victoria, BC, V9E 2E7, Canada}

\author{Helen Kirk}
\affiliation{NRC Herzberg Astronomy and Astrophysics, 5071 West Saanich Rd, Victoria, BC, V9E 2E7, Canada}

\author{Kevin Lacaille}
\affiliation{Department of Physics and Atmospheric Science, Dalhousie University, Halifax, NS, B3H 4R2, Canada}

\author{Geoffrey C. Bower}
\affiliation{Academia Sinica Institute of Astronomy and Astrophysics, 645 N. A`oh\={o}k\={u} Place, Hilo, HI 96720, USA}

\author{Graham S. Bell}
\affiliation{East Asian Observatory, 660 North A`oh\={o}k\={u} Place, University Park, Hilo, Hawaii 96720, USA}

\author{Sarah Graves}
\affiliation{East Asian Observatory, 660 North A`oh\={o}k\={u} Place, University Park, Hilo, Hawaii 96720, USA}

\author{Scott Chapman}
\affiliation{Department of Physics and Atmospheric Science, Dalhousie University, Halifax, NS, B3H 4R2, Canada}

\author{The JCMT Transient Team}
\altaffiliation{Yuri Aikawa, University of Tsukuba; Geoffrey Bower, ASIAA; Joanna Bulger, Subaru Telescope; Vivien Chen, National Tsing Hua University; Wen-Ping Chen, National Central University; Eun Jung Chung, KASI; Jennifer Hatchell, University of Exeter; Yuxin He, Xinjiang Astronomical Observatory; Gregory Herczeg KIAA/Peking University; Po-Chieh Huang, National Central University; Miju Kang, KASI; Sung-ju Kang, KASI; Gwanjeong Kim, KASI; Jongsoo Kim, KASI; Kyoung Hee Kim, KNU/KNUE; Mi-Ryang Kim, Chungbuk University; ShinYoung Kim, KASI/UST; Yi-Jehng Kuan, National Taiwan Normal University; Woojin Kwon, KASI/UST; Shih-Ping Lai, National Tsing Hua University; Bhavana Lalchand, National Central University; Chang Wong Lee, KASI; Jeong-Eun Lee, Kyung Hee University; Feng Long, KIAA/Peking University; A-Ran Lyo, KASI; Oscar Morata, ASIAA; Harriet Parsons, East Asian Observatory; Andy Pon, University of Western Ontario; Ramprasad Rao, ASIAA; Jonathan Rawlings, University College London; Manash Samal, National Central University; Aleks Scholz, St Andrews University; Peter Scicluna, ASIAA; Archana Soam, KASI; Dimitris Stamatellos, University of Central Lancashire; Wang Yiren, Peking University; Hyunju Yoo, Chungnam National University; Miaomiao Zhang, Max Planck Institute for Astrophysics; Jianjun Zhou, Xinjiang Astronomical Observatory}

\begin{abstract}

Though there has been a significant amount of work investigating the early stages of low-mass star formation in recent years, the evolution of the mass assembly rate onto the central protostar remains largely unconstrained. Examining in depth the variation in this rate is critical to understanding the physics of star formation. 
Instabilities in the outer and inner circumstellar disk can lead to episodic outbursts. Observing these brightness variations at infrared or submillimetre wavelengths sets constraints on the current accretion models. 
The JCMT Transient Survey is a three-year project dedicated to studying the continuum variability of deeply embedded protostars in eight nearby star-forming regions at a one month cadence. We use the SCUBA-2 instrument to simultaneously observe these regions at wavelengths of \mbox{450 $\mu$m} and \mbox{850 $\mu$m}. In this paper, we present the data reduction techniques, image alignment procedures, and relative flux calibration methods for \mbox{850 $\mu$m} data. We compare the properties and locations of bright, compact emission sources fitted with Gaussians over time. Doing so, we achieve a spatial alignment of better than 1$\arcsec$ between the repeated observations and an uncertainty of 2-3\% in the relative peak brightness of significant, localised emission. This combination of imaging performance is unprecedented in ground-based, single dish submillimetre observations. Finally, we identify a few sources that show possible and confirmed brightness variations. These sources will be closely monitored and presented in further detail in additional studies throughout the duration of the survey. 

\end{abstract}


\section{Introduction}
\label{introductionsec}

Although there have been many advances made in understanding low mass star formation over the past ten years (see, for example, \citealt{difrancesco2007}, \citealt{wardthompson2007review}, \citealt{andre2014}), the manner in which mass assembles onto a forming star remains a crucial open question. As \cite{kenyon1990} first demonstrated, assuming the mass accretion process onto a young star occurs at a constant rate (steady inside out collapse; \citealt{shu1987}) gives rise to ``The 
Luminosity Problem'': the empirical result that the median protostellar luminosity is measured to be approximately an order of magnitude less than the expected value. In recent years, this problem has been confirmed and emphasised by \textit{Spitzer Space Telescope} observations through which even lower luminosities have been discovered (see \citealt{dunham2008}, \citealt{evans2009}, \citealt{enoch2009}, \citealt{dunham2013}, \citealt{dunham2014}). One solution to this problem is that the accretion does not proceed at a constant rate. Rather, it occurs during episodic events which may be accompanied by outbursts that can be detected at infrared, submillimetre, and sometimes optical wavelengths (see \citealt{mckee2011}, \citealt{johnstone2013}, and \citealt{scholz2013}). There is indirect evidence that episodic accretion occurring while a protostar is still deeply embedded in its nascent gas and dust is an early phase of the more evolved FU Orionis (FUors; \citealt{herbig1977}, see also \citealt{hartmann1985}) sources \citep{dunham2014,audard2014}.

The physical mechanism responsible for a continuum outburst detected at the submillimetre wavelengths of interest to this survey is re-radiation from heated dust grains in the surrounding protostellar envelope. Outside of the JCMT Transient Survey, there have already been two millimetre sources (both embedded protostars) that have shown direct evidence of an active burst accretion phase accompanied by a dramatic brightening, 
HOPS 383 in Orion (\citealt{safron2015}; using 
 Atacama Pathfinder Experiment and SCUBA archive data), and MM1 in NGC6334I (\citealt{hunter2017}; using Atacama Large Millimeter/submillimeter Array and Submillimeter Array data). 
 
The JCMT Transient Survey (Herczeg et al. in preparation) is a three year project dedicated to observing continuum variability in deeply embedded protostars at submillimetre wavelengths with the Submillimetre Common User Bolometer Array 2 (SCUBA-2; \citealt{holland2013}). To this end, we are monitoring eight regions selected from the JCMT Gould Belt Survey (GBS; \citealt{wardthompson2007}) that have a high density of known protostellar and disk sources (Young Stellar Object Classes 0 to II and flat spectrum; see \citealt{lada1987}, \citealt{andre1993}, and \citealt{greene1994}) at an approximate 28 day cadence whenever they are observable in the sky. SCUBA-2 uses approximately 10,000 bolometers subdivided into two arrays to observe at both \mbox{450 $\mu$m} and \mbox{850 $\mu$m} simultaneously. While we expect sources undergoing an accretion burst event to show a stronger signal at \mbox{450 $\mu$m} \mbox{\citep{johnstone2013}}, in this paper we focus only on the \mbox{850 $\mu$m} data. The noise levels in \mbox{450 $\mu$m} maps are much more dependent on the weather than their \mbox{850 $\mu$m} counterparts, causing the signal-to-noise ratio (SNR) to fall dramatically when there is more water vapour in the atmosphere. In addition, the beam profile is less stable than at \mbox{850 $\mu$m} (as shorter wavelengths are more susceptible to dish deformation, and focus errors. For more information, see \citealt{dempsey2013}) requiring careful attention in order to make precise measurements of compact objects. We thus start here by defining the \mbox{850 $\mu$m} calibration and we will use this knowledge to calibrate the \mbox{450 $\mu$m} data at a later date. As the survey matures and precise \mbox{450 $\mu$m} data calibration is achieved, these simultaneous observations will provide further confirmation of significant variations.

In order to track the peak brightnesses of submillimetre emission sources over each epoch, we test and employ a robust data reduction method and use multiple observations of the same regions to derive post-reduction image alignment and relative flux calibration techniques.  Reducing SCUBA-2 data is a complex process with several user-defined parameters that affect the final image produced (for detailed information on SCUBA-2 data reduction procedures, see \citealt{chapin2013}). A large amount of work has been invested in understanding the optimal data reduction parameters to use for differing science goals (see, for example, \citealt{mairs2015}) depending on the scan pattern of the telescope and the amount of large-scale structure that needs to be robustly recovered. In all cases, the nominal JCMT pointing error is 2-6$\arcsec$ (East Asian Observatory staff, \textit{private communication}) and the flux calibration is uncertain to $\sim$5-10\% (\citealt{dempsey2013}; see also, Section \ref{fluxcal}). While this is sufficient for most projects which use JCMT data, both of these uncertainties can be improved upon when there are multiple observations of regions with bright sources taken in a consistent manner. In this work, we seek to improve both the spatial alignment and the flux calibration of the JCMT Transient Survey data by approaching the problem from a relative point of view. 

Properly matching faint, potential protostellar sources over the observed epochs and co-adding those observations with high precision for the highest SNR requires sub-pixel accuracy (<<3$\arcsec$ at \mbox{850 $\mu$m}) in the spatial alignment. Similarly, if we were to use the nominal flux calibration, where the uncertainty is taken to be $\sigma\sim10\%$, the flux would need to vary by 30-50\% for a transient event to be deemed significant (3-5$\sigma$). Thus, our goal is to reduce this uncertainty by a factor of 3 to 5 (i.e. sigma \mbox{$\sim 2-3\%$}) by considering relative brightness changes over time and ignoring the absolute flux calibration. We will then be able to measure flux variations of $\sim10\%$ to be statistically significant (>$3\sigma$). Several models predict smaller flux variations due to episodic accretion over few year timescales to occur much more frequently than large flux variations (see, for examples, \citealt{bae2014},  \citealt{vorobyov2015}, and Herczeg et al. in preparation). Observations like those performed throughout the JCMT Transient Survey will help constrain the current models. The techniques we have developed here can be applied to any JCMT data obtained in a similar manner, including archival data obtained by the GBS (follow up analysis by Mairs et al., in prep.). Thus, we are able to successfully align and relatively flux calibrate archival data such as those which were obtained by the GBS and we include these data in a follow-up analysis (Mairs et al. in preparation).

This paper is organised as follows: In Section \ref{obssec} we summarise the details of our SCUBA-2 observations. In Section \ref{drsec} we outline our data reduction methods and showcase four tests we performed which altered the amount of large-scale structure recovered in a given map and the initial priors offered to the map-making pipeline in order to select the most robust techniques for our purpose of detecting transient events in deeply embedded protostars. In Section \ref{postredcal} we detail our source extraction, post-reduction spatial alignment, and relative flux calibration methods applied to all current JCMT Transient data. In Section \ref{discusssec}, we present an analysis on the recovered compact emission sources and highlight objects of interest including the first demonstrably variable source in our survey (Yoo et al. in preparation).
Finally, we present our conclusions in Section \ref{conclusionsec}.

\vspace{5mm}

\section{Observations}
\label{obssec}

\begin{table*}
\caption{A summary of the observed JCMT Transient Survey fields between the first observations on December 22$^\text{nd}$, 2015 and March 1$^\text{st}$, 2017.}
\label{regionlist}
\begin{tabular}{|c|c|c|c|c|c|c|}
\hline
Region & \multicolumn{1}{|p{1.5cm}|}{\centering Central R.A. \\ (J2000)} & \multicolumn{1}{|p{1.5cm}|}{\centering Central Decl. \\ (J2000)} & \multicolumn{1}{|p{1.2cm}|}{\centering Number of\\ Epochs$^{a}$} & \multicolumn{1}{|p{2.25cm}|}{\centering Average \mbox{850 $\mu$m} Noise$^{b,c}$\\ (mJy\: beam$^{-1}$)} & \multicolumn{1}{|p{2.25cm}|}{\centering Std. Dev. \mbox{850 $\mu$m} Noise$^{d}$\\ (mJy\: beam$^{-1}$)} & \multicolumn{1}{|p{2.1cm}|}{\centering Noise in the Co-add\\ (mJy\: beam$^{-1}$)}\\ 
\hline\hline
Perseus: NGC1333 & 03:28:54 & +31:16:52 & 10 & 12.26 & 0.40 & 3.92\\
Perseus: IC348 & 03:44:18 & +32:04:59 & 9 & 12.18 & 0.43 & 4.30\\
Orion: OMC 2/3 & 05:35:33 & -05:00:32 & 9 & 11.72 & 0.54 & 4.19\\
Orion: NGC2024 & 05:41:41 & -01:53:51 & 11 & 11.29 & 0.40 & 4.32\\
Orion: NGC2068 & 05:46:13 & -00:06:05 & 10 & 11.75 & 0.38 & 3.85\\
Ophiuchus: Core & 16:27:05 & -24:32:37 & 8 & 13.35 & 0.75 & 5.00\\
Serpens: Main & 18:29:49 & +01:15:20 & 9 & 12.01 & 0.27 & 4.54\\
Serpens: South & 18:30:02 & -02:02:48 & 9 & 14.56 & 1.18 & 4.72\\
\hline
\end{tabular}
\begin{flushleft}
$^{a}$ Only observations between December 22$^\text{nd}$, 2015 and March 1$^\text{st}$, 2017 are included.\\ 
$^{b}$These measurements of the \mbox{850 $\mu$m} noise levels are based on a point source detection in a single observation using 3$\arcsec$ pixels and a 14.6$\arcsec$ FWHM beam.\\
$^{c}$The reduction method \textit{R3} was used to derive these noise estimates (see Section \ref{drsec}).\\
$^{d}$The standard deviation of the average \mbox{850 $\mu$m} noise across all epochs.
\end{flushleft}
\end{table*}

 The JCMT Transient Survey observations are performed simultaneously at 450 and \mbox{850 $\mu$m} with effective beam sizes of 9.8\arcsec and 14.6\arcsec \citep{dempsey2013}, respectively, using the Submillimetre Common User Bolometer Array 2 (SCUBA-2; \citealt{holland2013}). We use the \textit{pong 1800"} mapping mode \citep{kackley2010} which yields circular maps 0.5$^{\circ}$ in diameter called ``pongs''. Each pixel (3$\arcsec$ at \mbox{850 $\mu$m}, 2$\arcsec$ at \mbox{450 $\mu$m}) contains the signal from several bolometers as the telescope scans across the sky, changing direction when it reaches the boundary of the circular field. This scan pattern ensures that each part of the field is observed from multiple position angles, resulting in the recovery of real astronomical structure while short timescale variations due to the atmosphere are attenuated. Eight nearby (\mbox{<500 pc}) regions selected from the Gould Belt Survey (\citealt{wardthompson2007}, Herczeg et al. in preparation) are each monitored at an approximate 28 day cadence whenever they are observable in the sky. Contained within these regions are a total of 1749 young stellar objects (YSOs) identified by \textit{Spitzer Space Telescope} (\citealt{megeath2012}, \citealt{dunham2015}) and \textit{Herschel Space Observatory} \citep{stutz2013} observations. 344 of these YSOs are identified as Class 0/I or Flat Spectrum protostars while the remaining 1405 are identified as Class II (disk) sources (Herczeg et al. in preparation). Table \ref{regionlist} shows a list of the regions and their central coordinates along with the average \mbox{850 $\mu$m} rms noise measured in the individual maps (see Appendix \ref{obsappendix} for detailed information on each individual observation). Note that Serpens South has diffuse structure throughout the map even near the edges of the field and, as a result, the measured noise is slightly higher than the other regions. There are five weather grades defined for JCMT observations from Band 1 (very dry: $\tau_{225\mathrm{\:GHz}}<0.05$, where $\tau_{225\mathrm{\:GHz}}$ is the zenith opacity of the atmosphere at \mbox{225 GHz}) to Band 5 (very wet: $\tau_{225\mathrm{\:GHz}}>0.2$). All of the observations performed in this survey were taken in either Band 1, Band 2, or Band 3 weather ($\tau_{225\mathrm{\:GHz}}<0.12$) as measured by the JCMT Water Vapour Radiometer \cite{dempsey2008}. The observing time per observation is set to 20-40 min, depending on the weather band, to maintain a similar noise quality of \mbox{$\sim 10\mathrm{\:mJy\:beam}^{-1}$} at \mbox{850 $\mu$m} (see Table \ref{regionlist}).

 Due to the higher telluric absorption (see \citealt{dempsey2013}) and varying PWV (precipitable water vapour), the \mbox{450 $\mu$m} observations have noise values 10 to 40 times higher than the \mbox{850 $\mu$m} observations and, thus, throughout this paper we focus on the latter. Observations began in December 2015 and are expected to continue until January 2019. Here, we present results between the first observations in December, 2015 and March 1$^\text{st}$, 2017.

\begin{figure*}
\centering
\subfloat{\label{}\includegraphics[width=8.5cm,height=7.8cm]{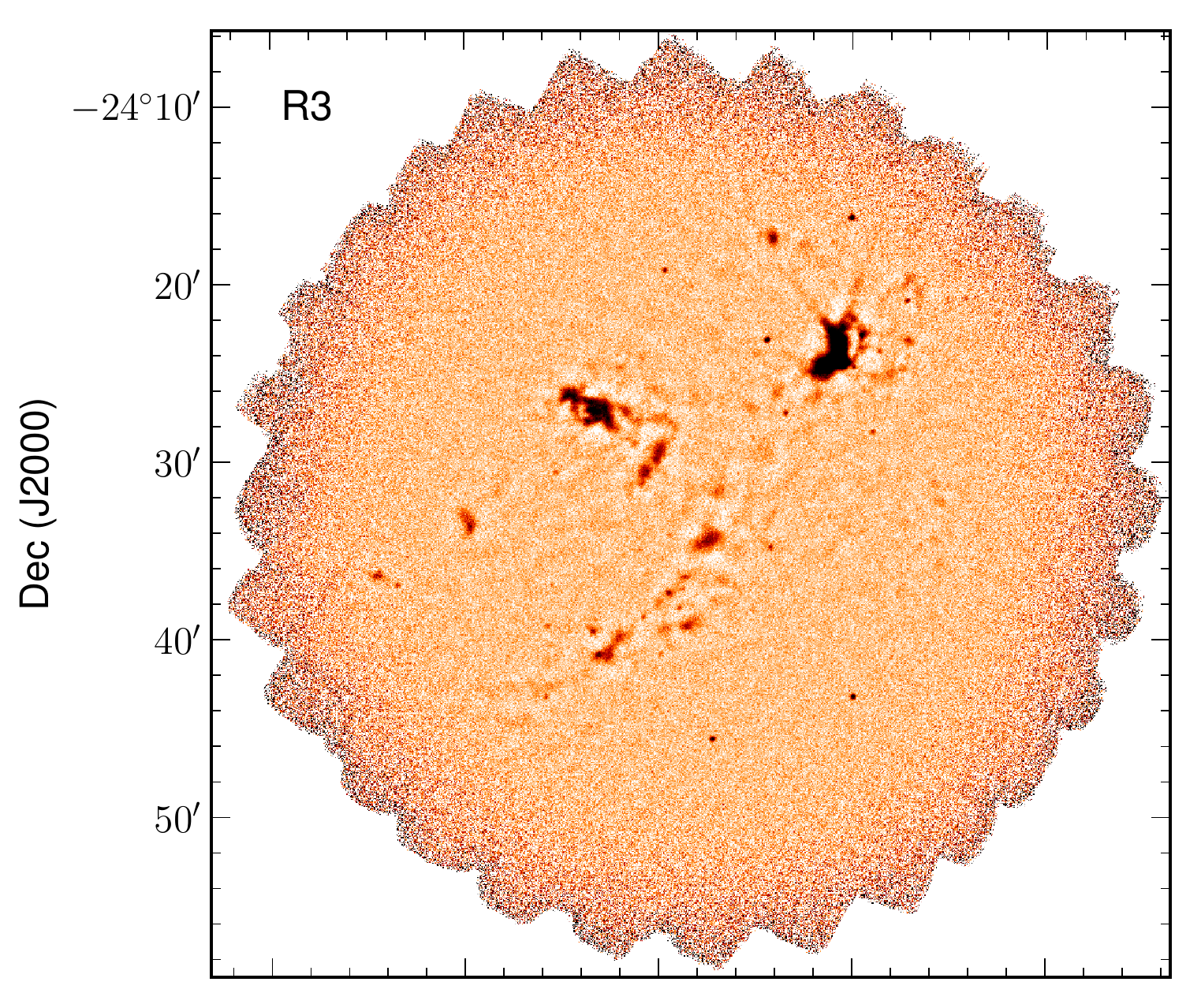}}
\subfloat{\label{}\includegraphics[width=8.5cm,height=7.8cm]{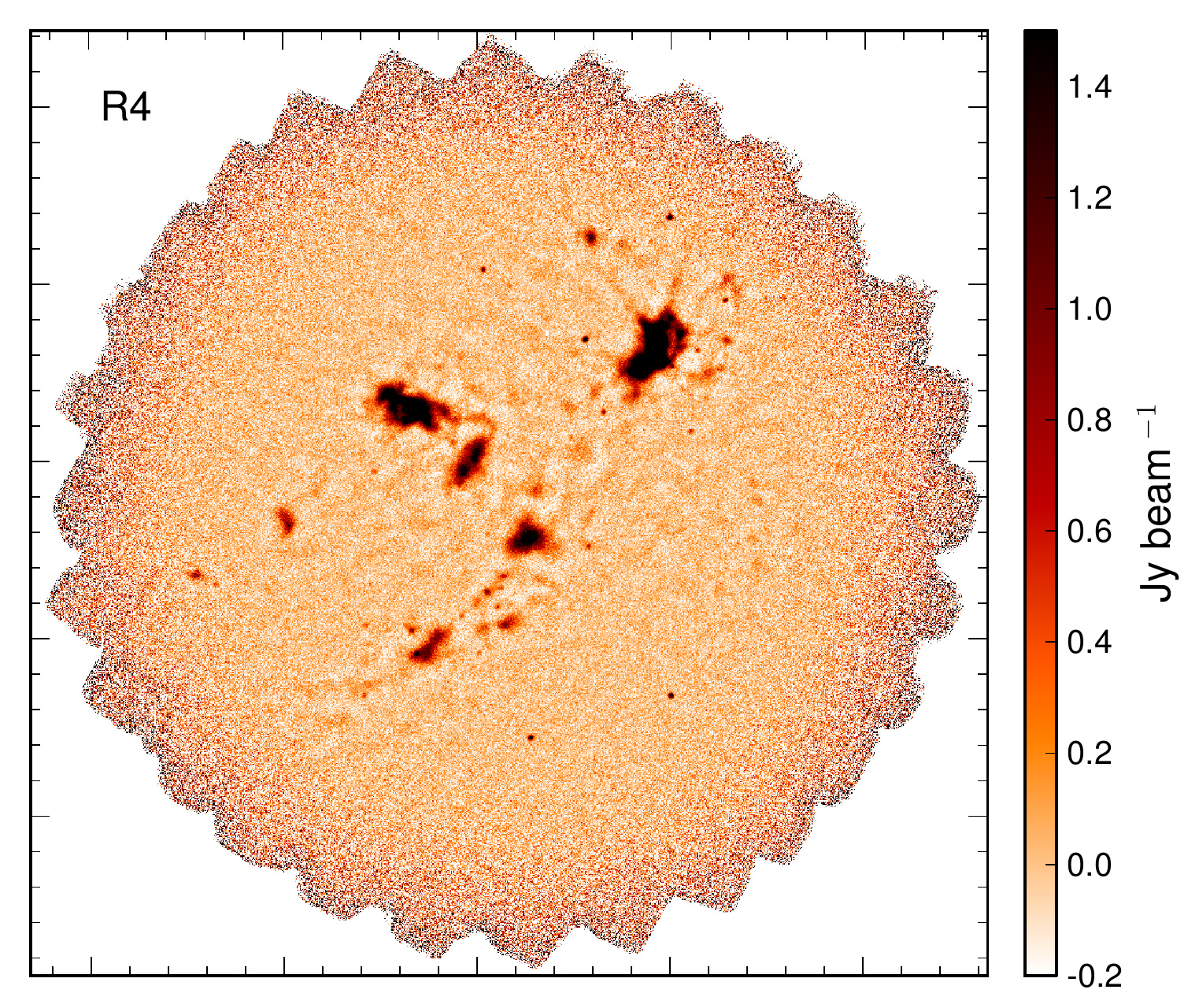}}\\
\subfloat{\label{}\includegraphics[width=8.5cm,height=7.8cm]{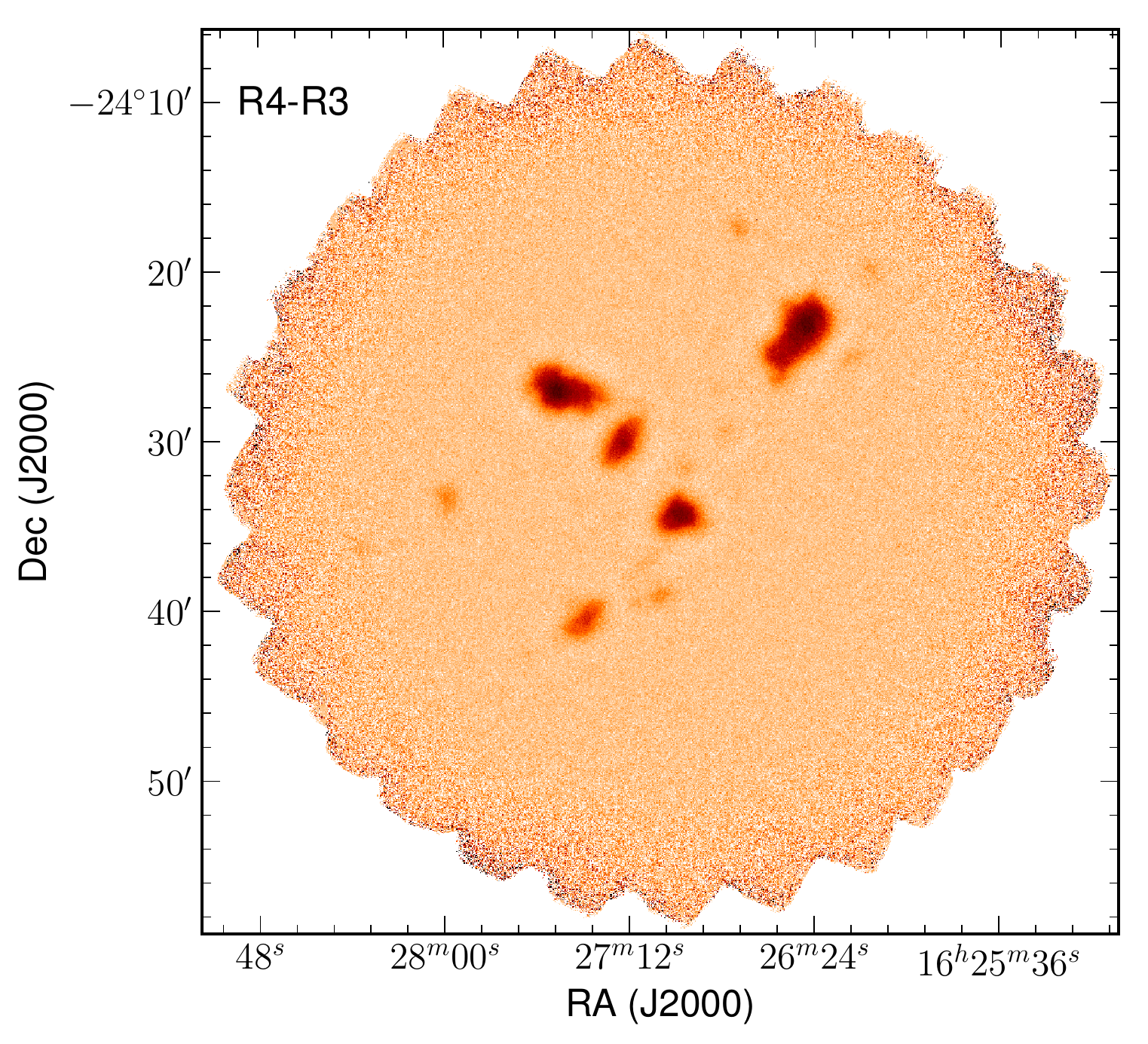}}
\subfloat{\label{}\includegraphics[width=8.5cm,height=7.8cm]{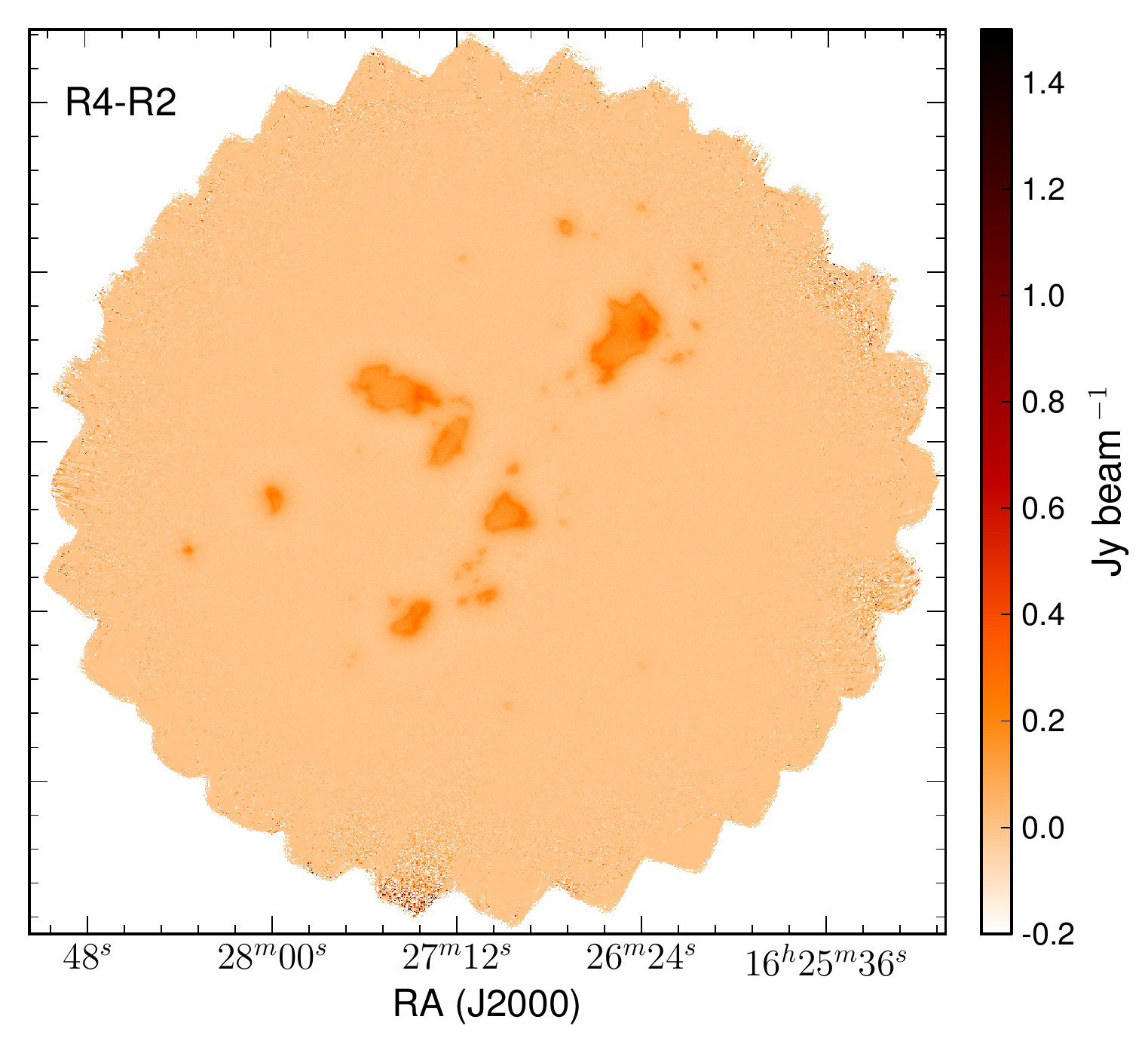}}\\
\caption{\textit{Top}: A single \mbox{850 $\mu$m} observation of the Ophiuchus Core region reduced using reduction methods \textit{R3} (left) and \textit{R4} (right). \textit{Bottom}: Reduction \textit{R4} minus reduction \textit{R3} (left) and reduction \textit{R4} minus reduction \textit{R2} (right). Note that the two small bright sources seen in bottom of the R3 and R4 maps are roughly point-like.}
\label{subfig}
\end{figure*}

\section{Data Reduction Methods}
\label{drsec}

The data reduction procedure was performed using the iterative map-making technique {\sc{makemap}} (explained in detail by \citealt{chapin2013}) in the {\sc{SMURF}} package (\citealt{jenness2013}) found within the {\sc{Starlink}} software \citep{currie2014}. Briefly, {\sc{makemap}} begins with the raw power detected by the telescope as a function of time throughout the duration of the observation and iteratively works to recover the astronomical signal by modeling and subtracting different sources of noise. First, general fixes (such as removing noise spikes, ensuring continuity, and apodising the edges of the bolometer time series) and a flat field correction are applied. Then, the program removes the \textit{common mode} (COM) noise. This type of noise, the large majority of which is caused by atmospheric emission, dominates the astronomical features we seek to study by causing a significant fraction of SCUBA-2's bolometers to acquire the same signal. As SCUBA-2 scans across a region of the sky at a constant speed, the power received over a given time interval is directly correlated with a spatial scale. Thus, removing the common mode noise results in a loss of extended, faint structure in the final maps produced, while the more compact, bright structure can be more accurately recovered. For an overview of the Gould Belt Survey for their Legacy Release 1 (GBS LR1) filtering parameters as well as results from testing the completeness of this method using artificial sources, see \cite{mairs2015}. 

Next, an atmospheric extinction model is applied based on the amount of PWV which was measured during the observation and a second filtering of the data is applied to remove any residual, low frequency 1/\textit{f} noise which was missed by the common mode subtraction. The extent of the final spatial filtering executed in this step is defined by the user. Generally, the largest recoverable scales are $\sim600\arcsec$ before atmospheric signal becomes significant. 

Finally, the astronomical signal is estimated and the residual white noise is compared to the previous iteration. The iterative solution converges when the difference in individual pixels changes on average by 
a user-defined threshold percentage of the rms noise present in the map (we select a value of $<0.1\%$). The maps produced are originally in units of picowatts (pW) but are converted to \mbox{mJy arcsec$^{-2}$} using the standard \mbox{850 $\mu$m} aperture flux conversion factor \mbox{2340 mJy pW$^{-1}$ arcsec$^{-2}$} and \mbox{4710 mJy pW$^{-1}$ arcsec$^{-2}$} at \mbox{450 $\mu$m} \citep{dempsey2013}. In the case of the JCMT Transient Survey, the final maps are gridded to 3$\arcsec$ pixels for the \mbox{850 $\mu$m} data and 2$\arcsec$ pixels for the \mbox{450 $\mu$m} data.

To apply additional constraints to the solution derived by {\sc{makemap}}, the user can also supply an \textit{external mask} which surrounds the astronomical signal deemed to be significant. To construct an external mask, the individual maps produced by the iterative mapmaker are co-added in order to achieve a higher signal-to-noise ratio (SNR). The resulting image is used to define the regions of genuine astronomical emission (pixels with a signal-to-noise ratio of at least 3 are generally deemed significant). This mask is then used in a second round of data reduction in order to recover better any faint and extended structure.   

{\sc{makemap}} has over 100 user defined parameters that allow full control over each step of the iterative process. Many parameters will cause negligible changes in the final maps produced, but some will cause significant differences, such as the extent of spatial filtering the user applies. For the JCMT Transient Survey, we begin with the robust data reduction parameters derived by the GBS LR1 dataset as described in \cite{mairs2015} 
To ensure, however, that we produced the best calibrated maps which would allow for the detection of the variability of embedded protostars, we tested the effects of altering the size of the spatial filter applied to the data to determine whether or not it was beneficial to apply additional constraints to {\sc{makemap}} by using an external mask.

 To this end, we performed four individual data reductions labeled \textit{R1}, \textit{R2}, \textit{R3}, and \textit{R4}, yielding four sets of maps exhibiting different models of the recovered, astronomical structure. 
 
 \begin{enumerate}[labelindent=0pt,labelwidth=\widthof{\ref{last-item}},label=\arabic*.,itemindent=1em,leftmargin=!]

\item \textit{R1}: An effective spatial filter of 200$\arcsec$ was applied and no external mask was used. 

\vspace{3mm}

\item \textit{R2}: An effective spatial filter of 600$\arcsec$ was applied and no external mask was used.

\vspace{3mm}

\item \textit{R3}: An external mask was constructed from a co-add of the \textit{R1} reduction and was used to constrain the solution derived by {\sc{makemap}}. Thus, the structure was filtered to 200$\arcsec$.

\vspace{3mm}

\item \textit{R4}: Similarly to \textit{R3}, an external mask was constructed from a co-add of the \textit{R2} reduction and applied to the data. Thus, the structure was filtered to 600$\arcsec$. This reduction is the same as the GBS LR1 data release \citep{mairs2015}.

\end{enumerate}

Figure \ref{subfig} shows an example of a single observation of the Ophiuchus Core region (see Table \ref{regionlist}). The top two panels show the resulting maps produced by reductions \textit{R3} (left) and \textit{R4} (right) while the bottom two panels show the subtraction of the \textit{R3} image from the \textit{R4} image (left) to highlight the effect of changing the effective spatial filter and the subtraction of the \textit{R2} image from the \textit{R4} image (right) to highlight the effect of the external mask. Though more extended structure is present in the \textit{R2} and \textit{R4} maps, the compact structure is recovered whether a 200$\arcsec$ or a 600$\arcsec$ filter is used. For the extended emission reconstructions, the mask returns additional flux, some of which appears pedestal-like. As \cite{mairs2015} discuss, the amount of extended structure which is recovered can produce slightly different results for the compact structure present in the map as the larger-scale background may add a pedestal value to the flux. 
We minimise the effect of the pedestal by using a localised peak extraction algorithm (described in Section \ref{postredcal} and Appendix \ref{gcappendix}) that filters out the extended background. 

 The dominant uncertainty between these different reduction methods is how the recovered extended structure and external masks alter the fit to compact emission sources. The measurement of the peak brightness of a source relies on a consistent procedure from observation to observation in conjunction with the optimal data reduction method. By fitting Gaussian profiles to compact emission sources and comparing their centroid positions and peak brightnesses (described in detail in Section \ref{postredcal}), we determine reduction \textit{R3} to be the most stable for our purposes (although all four reductions work reasonably well). The external mask limits the flux distribution during the map-making procedure while the harsher filter (200$\arcsec$ as opposed to 600$\arcsec$) subdues large-scale structure which is not expected to vary (but is hard to recover due to signal from the atmosphere and the instrumentation).

SCUBA-2's \mbox{850 $\mu$m} filter coincides with the broad $^{12}$CO(J=3-2) emission line. No attempt has been made to remove this signal. This excess flux, however, will not affect our ability to measure precisely the variability of deeply embedded protostars. As \cite{drabek2012} and \cite{coude2016} discuss, the CO(J=3-2) line contributes only low-level emission ($\leq$ 20\%) except for a few sources of stellar outflow. In addition, as \cite{mairs2016} show, the peak brightnesses of compact sources are not affected by the removal of the emission line. 

\section{Post Reduction Calibrations}
\label{postredcal}

Since the JCMT Transient Survey is interested in measuring the fluxes of individual compact sources over time, it is important to take into consideration both the pointing uncertainty of the telescope (expected to be 2 to 6 arcseconds; East Asian Observatory staff \textit{private communication}) as well as the flux calibration uncertainty (expected to be $\sim5-10\%$, 
see \citealt{dempsey2013}). To this end, we perform two post-reduction calibrations: 1) We derive and apply a pointing correction to more precisely align the maps with one another and 2) We derive and apply a relative flux calibration factor for each image produced in order to consistently compare a given source from observation to observation. Since both of these calibrations are relative corrections for each region, we can use the most robust, compact emission sources present in each map to calibrate self-consistently. The first step is to identify the appropriate calibrator sources in each of the eight regions. 

There are many different, publicly available algorithms designed to extract structure from a given region (for examples, see {\sc{Gaussclumps}} \citealt{stutzki1990}, {\sc{ClumpFind}} \citealt{williams1994}, 
{\sc{Astrodendro}} \citealt{rosolowsky2008}, {\sc{getsources}} \citealt{getsources2012}, 
 and {\sc{FellWalker}} \citealt{berry2015}).  Each method combines detected emission differently based on user supplied criteria and, thus, the use of such algorithms requires discernment and a goal-based approach. In this work, we are interested in accurately determining the brightness of localised, compact sources in dust emission convolved with the JCMT beam, which we expect to have approximately Gaussian features. The most robust (often isolated) Gaussian sources will be used for image calibration. To this end, we have selected the algorithm {\sc{Gaussclumps}} \citep{stutzki1990} to identify and extract sources in each observation of a given field as this program is designed to robustly characterise Gaussian structure and subtract background structure, such as pedestals. Specifically, we use the {\sc{starlink}} software \citep{currie2014} implementation of {\sc{Gaussclumps}} found within the {\sc{cupid}} \citep{berry2007} package. For more information on {\sc{Gaussclumps}}, refer to Appendix \ref{gcappendix}.

\subsection{Image Alignment}
\label{imagealign}

To perform the post-reduction relative image alignment, we focus on the \mbox{850 $\mu$m} data. The noise in this dataset is measured to be more than an order of magnitude below its \mbox{450 $\mu$m} counterpart (due to the effect of PWV in the atmosphere) and the beam profile has greater stability, allowing us to more reliably fit the compact emission sources. The \mbox{450 $\mu$m} and \mbox{850 $\mu$m} data are, however, taken simultaneously, so the same alignment correction is applied to both datasets. The alignment procedure we apply to the data consists of five steps:

\begin{figure*}	
\centering
\includegraphics[width=16.7cm,height=7.7cm]{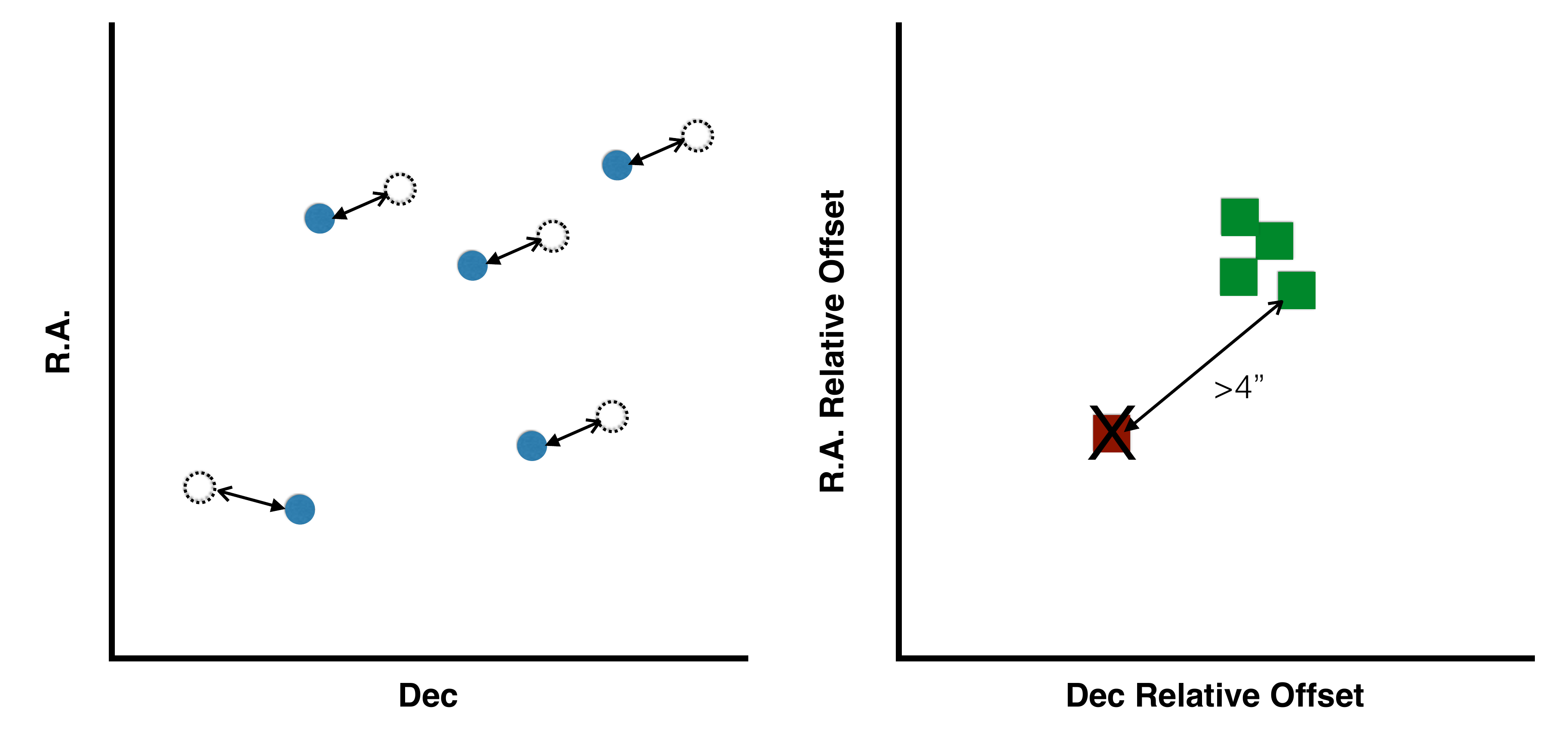}
\caption{Schematic diagram of step 4 in the image alignment process. \textit{Left:} We measure the offsets between bright, compact sources in the reference map (filled circles) and in a subsequent observation (empty circles). \textit{Right:} We compare the relative right ascension and declination offsets of all the sources and remove outliers.}
\label{offsetcull}
\end{figure*}

\begin{enumerate}[labelindent=0pt,labelwidth=\widthof{\ref{last-item}},label=\arabic*.,itemindent=1em,leftmargin=!]

\item We label the first observation of a given field the ``reference observation''. Then, we smooth the map with a 6$\arcsec$ Gaussian kernel to mitigate noise fluctuations. We identify and fit Gaussians to all the sources brighter than \mbox{200 mJy\: beam$^{-1}$} and with radii less than 10$\arcsec$ in the reference observation using the source extraction algorithm {\sc{Gaussclumps}} \citep{stutzki1990}. The radius of a source, $r$, is defined as $r = \sqrt{\mathrm{FWHM_{1}}\times \mathrm{FWHM}_{2}}/2$ where the FWHM$_{N}$ terms are the full widths at half maximum of the fitted two dimensional Gaussian. Fitting Gaussians to bright sources allows us to measure the centroid location of the sources to sub-pixel accuracy. For more information on how {\sc{Gaussclumps}} was executed, see Appendix \ref{gcappendix}.

\vspace{3mm}

\item For later observations, we also smooth the maps and identify and fit Gaussians to all the sources brighter than \mbox{200 mJy\: beam$^{-1}$} and with radii less than 10$\arcsec$ using {\sc{Gaussclumps}} in the same manner as for the reference observation. 

\vspace{3mm}

\item We next match each source identified in the reference catalogue to the nearest source in the later catalogue (the peak location sources must not differ by more than 10$\arcsec$, given an expectation that the alignment offset is better than this value). 

\vspace{3mm}

\item We then perform a check to ensure that we have matched the reference sources to the correct corresponding sources in the later catalogue by employing a simple test. Plotting the relative right ascension offset against the relative declination offset for all of the sources, we search for outliers by applying the condition that the resultant offset of every source must be within $4\arcsec$ of the resultant offset of any other source (see Figure \ref{offsetcull}). $4\arcsec$ was chosen after extensive testing across all eight regions revealed this threshold to consistently eliminate outliers. If this condition fails, we exclude that source from the final step. In this way, any moving sources or spurious detections will be discounted from our analysis. 

\vspace{3mm}

\item Finally, we average the right ascension and declination offsets of the matched sources to find the difference between the position of the later observation and the reference observation. We then apply this offset by re-reducing the later observation and correcting the pointing using {\sc{makemap}}'s \textit{pointing} parameter. In this way, both the reference and the later observations will be consistently aligned, processed, and gridded to the same world coordinate system grid.

\end{enumerate}

\begin{figure} 	
\centering
\includegraphics[width=9cm,height=6.7cm]{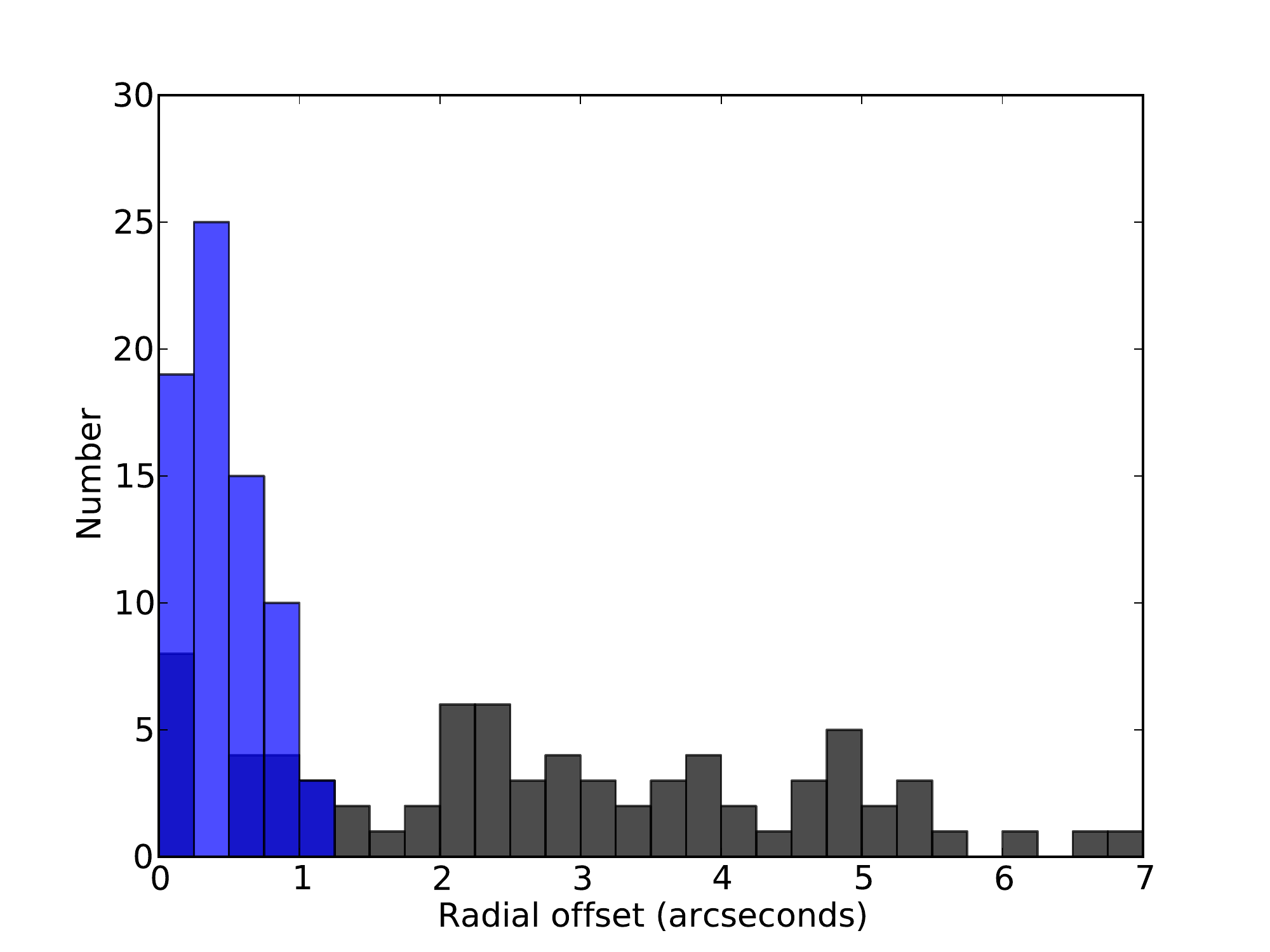}
\caption{Histograms of the measured radial offset between each region's reference field and its subsequent observations. Black represents the original offset without applying any correction; blue represents the corrected offset of the aligned maps.}
\label{align_fig}
\end{figure}

\begin{figure}
\centering
\includegraphics[width=0.5\textwidth]{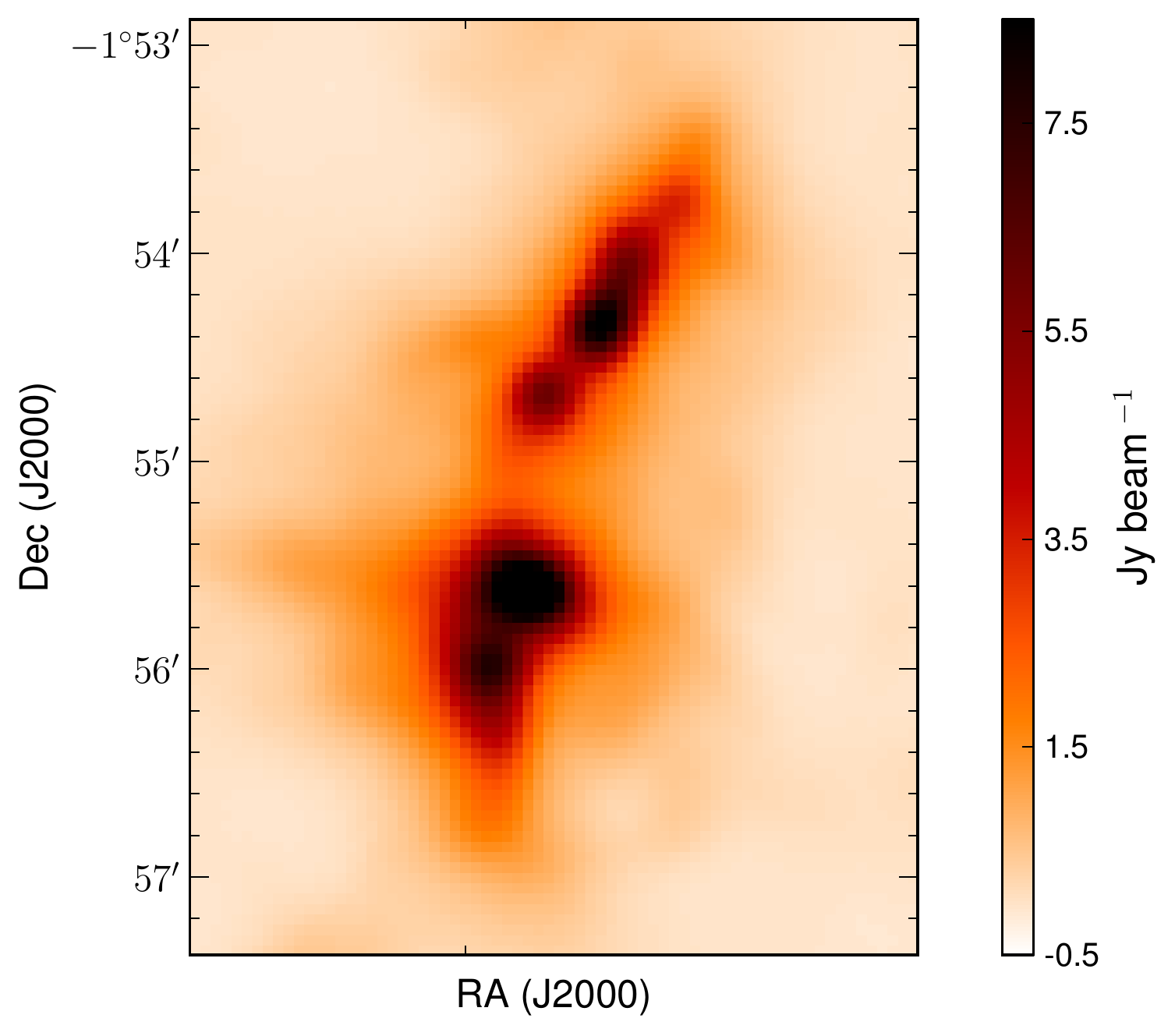}
\caption{The central region of NGC2024 at \mbox{850 $\mu$m}. The tightly clustered sources of emission are blended, causing a higher uncertainty in the Gaussian fits of the individual peaks.}
\label{ngc2024cutout}
\end{figure}

Figure \ref{align_fig} shows the 
results of applying the post-reduction alignment to the JCMT Transient Survey observations using reduction \textit{R3} (all four reduction methods show consistent results, see Appendix \ref{obsappendix}). The black histogram shows the original pointing uncertainty 
while the blue histogram shows the corrected pointing uncertainty. By reliably fitting bright peaks and matching their centroids from observation to observation, we have achieved a mean positional uncertainty of $0.45\arcsec$ (less than one sixth the width of an \mbox{850 $\mu$m} pixel) with a standard deviation of 0.3$\arcsec$. In nearly all cases, the images are aligned to better than $1\arcsec$. The few fields that exhibit a slightly higher uncertainty come from the NGC2024 region which contains more clustered sources mixed with larger-scale structure (see Figure \ref{ngc2024cutout} for an example of the clustered emission in NGC2024 and see Herczeg et al. in preparation for co-added images of all eight fields). Isolated, bright emission sources have less fitting uncertainties and therefore produce the best alignments. The alignment of the maps is now part of an automated routine run at the East Asian Observatory (EAO) immediately after the observations are taken at the telescope. The final, aligned images and {\sc{Gaussclumps}} catalogues are deposited in a shared directory which team members can access. 

In addition to this image alignment procedure, an independent method based on the cross-correlation of the observations was also tested and found to produce consistent results (see Appendix \ref{kevinappendix}). As the survey matures, we will be exploring this alternate technique and refining our methodology to further improve our alignment calibration. 

\subsection{Relative Flux Calibration}
\label{fluxcal}

\begin{figure} 	
\centering
\includegraphics[width=8.5cm,height=7cm]{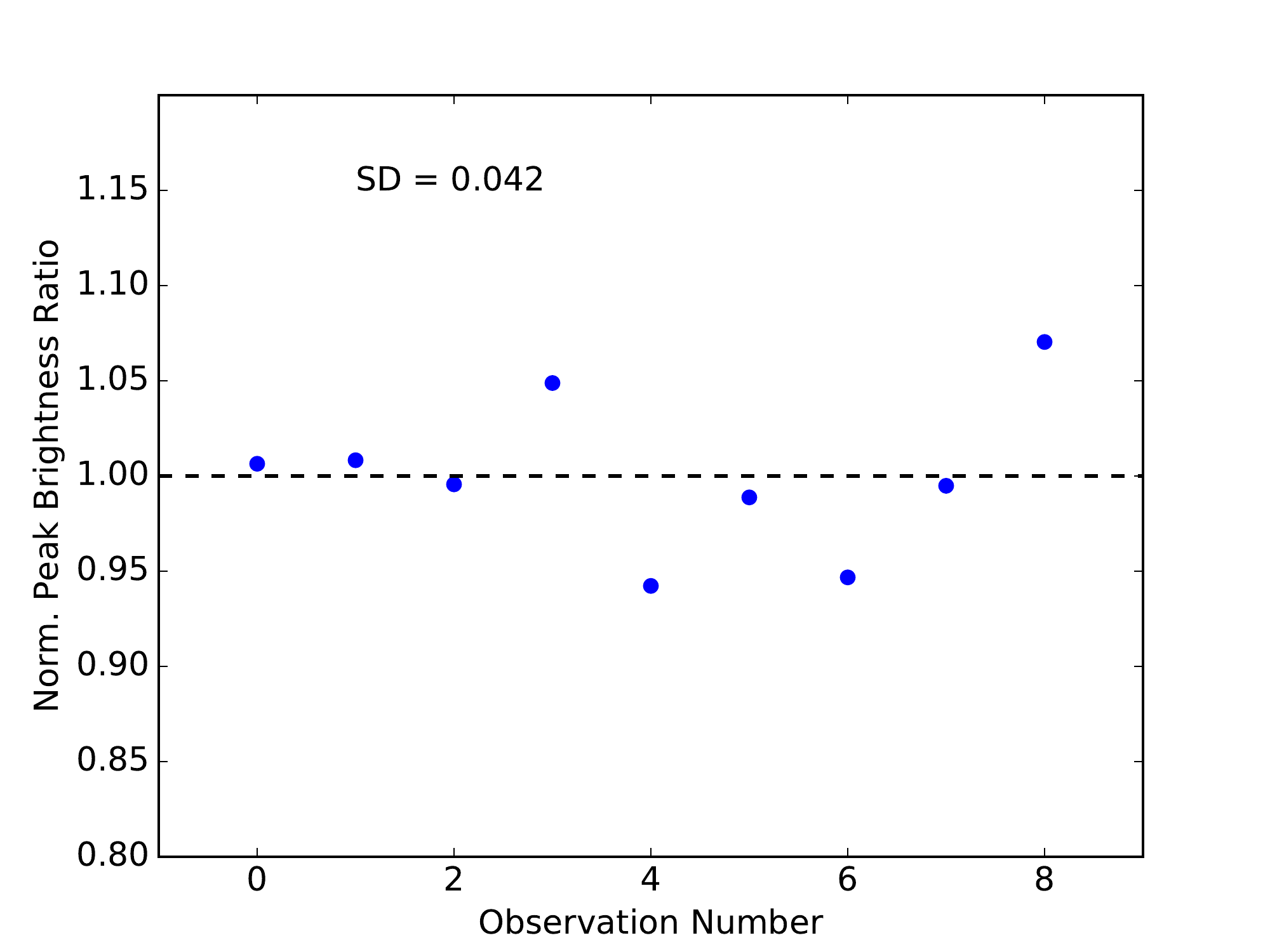}
\caption{An example of the normalised peak brightness of one source divided by the normalised peak brightness of another, plotted from observation to observation. 
SD is the standard deviation of this set of nine points, highlighting that the uncertainty in the ratio of these two sources (and the underlying uncertainty in the measurement of these individual sources) is about 4\%.}
\label{Rfig}
\end{figure}

\begin{figure*} 	
\centering
\includegraphics[width=12cm,height=10cm]{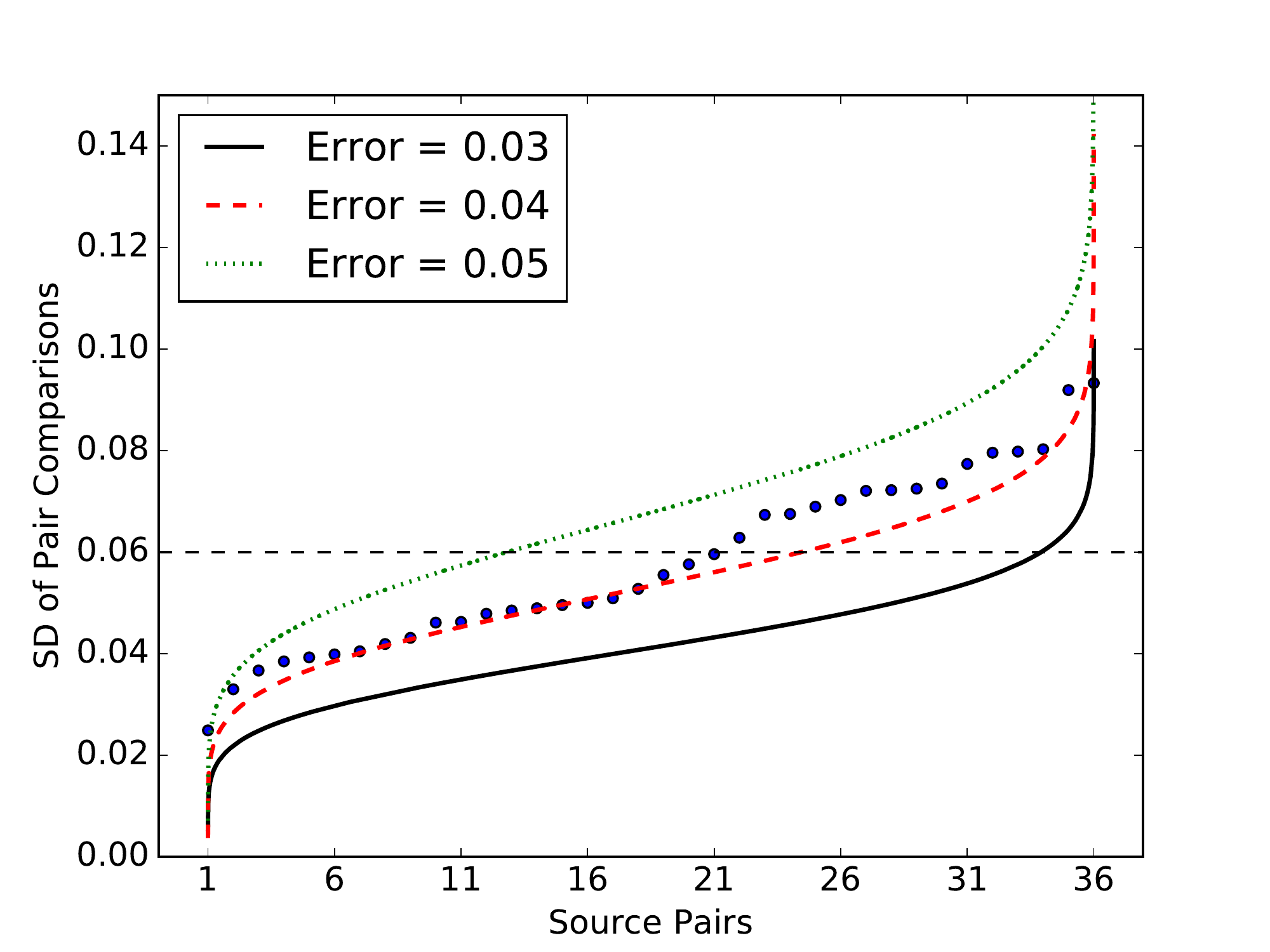}
\caption{The standard deviation of the normalised peak brightness ratios for all pairs of identified sources including all 8 observations of the Ophiuchus Core region observed prior to March 1$^\text{st}$, 2017, arranged in ascending order. Each line represents an iteration of a simple model where we applied a Gaussian error of the value indicated to 1000 sources of peak brightness 1.0 and compared their normalised peak brightness ratios over 8 epochs. Note that as more observations are performed, the central part of the curve flattens, approaching a value of $\sqrt{2}\times\mathrm{\:error}$. Nine potential calibrator sources were found, yielding 36 pairs. The largest \textit{Family} of sources consistent with one another (standard deviations less than 0.06, the threshold indicated by the dashed black line) are the flux calibrator sources we select to perform the correction. In this case, four sources met the criteria to join the flux calibrator \textit{Family}.}
\label{stdevplot}
\end{figure*}

\begin{figure*}
\centering
\subfloat{\label{}\includegraphics[width=9cm,height=7.8cm]{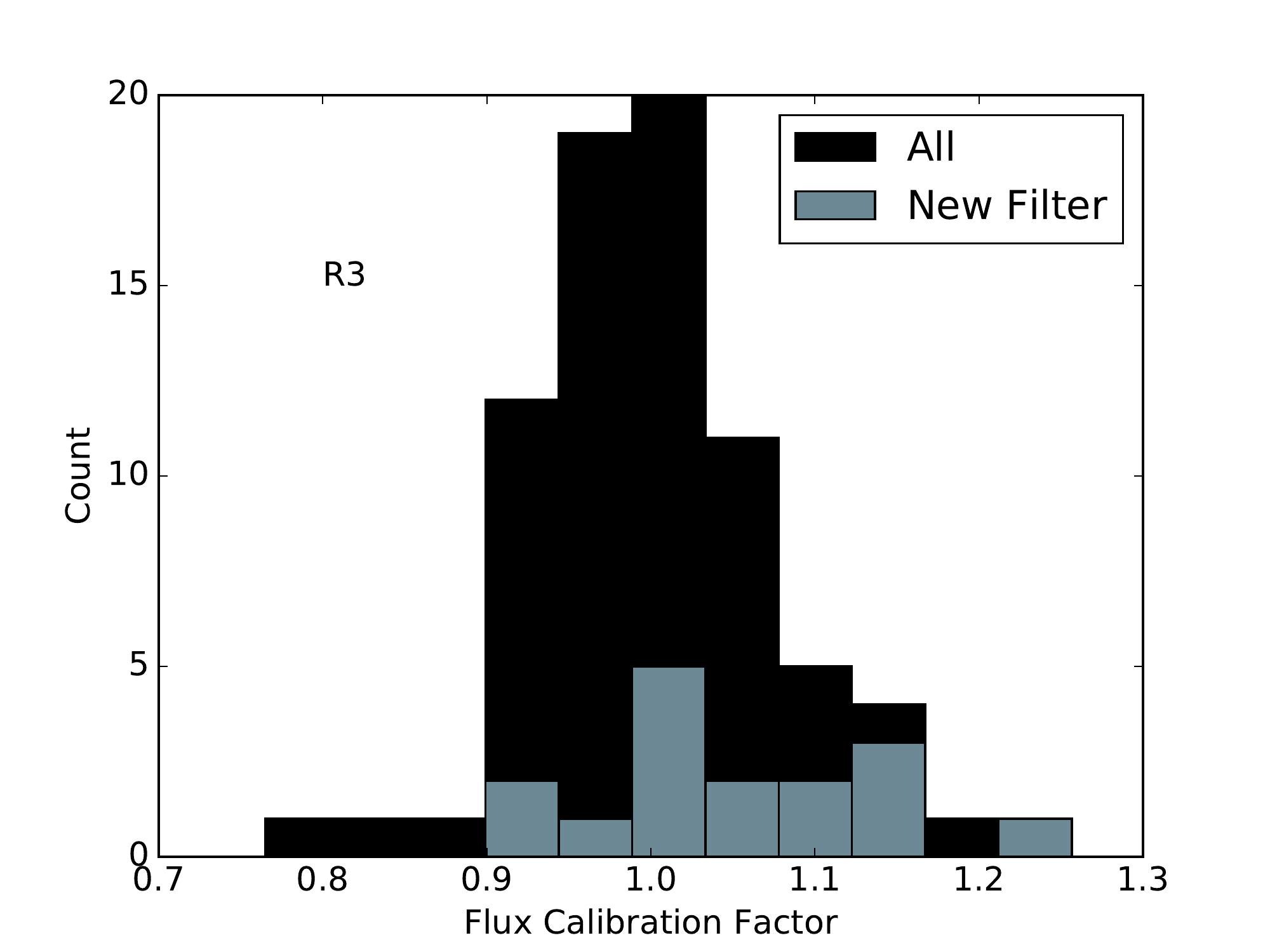}}
\subfloat{\label{}\includegraphics[width=9cm,height=7.8cm]{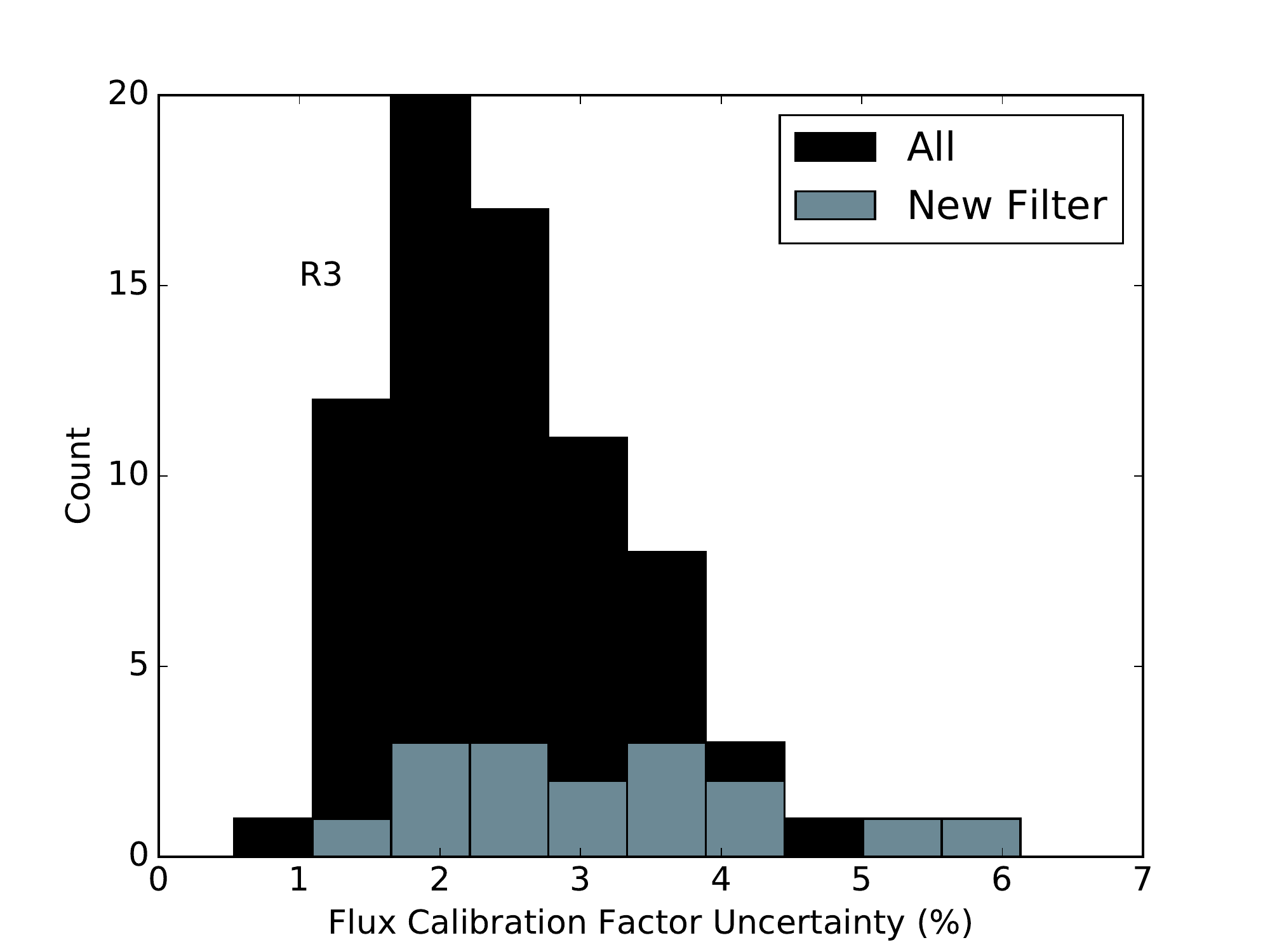}}
\caption{\textit{Left:} The derived flux calibration factors for compact emission sources for all observations of the eight regions. \textit{Right:} The relative uncertainty in the flux calibration factors, calculated by finding the standard deviation of the normalised peak brightnesses of the calibrator sources in each respective image. Black indicates all observations taken before March 1$^\text{st}$, 2017 while grey indicates observations take after the filter change of November, 2016.}
\label{fcf_fig}
\end{figure*}

After cataloguing the sources in the reference observation as well as the subsequent, aligned observations, we derive and apply a relative flux calibration factor to each observation in order to plot accurately the brightness variations of a given object over all epochs. The JCMT has an intrinsic absolute flux calibration of $\sim$\mbox{5-10\%} \citep{dempsey2013}, but we are focused only on the relative brightness changes from epoch to epoch. This allows us to achieve more accurate measurements of the variability in a given field. The procedure to flux calibrate our images consists of six steps:

\begin{enumerate}[labelindent=0pt,labelwidth=\widthof{\ref{last-item}},label=\arabic*.,itemindent=1em,leftmargin=!]

\item Beginning with the same calibrator sources that we extracted to perform the spatial alignment of the maps (described in Section \ref{imagealign}), we select the subset that have peak brightnesses over \mbox{500 mJy beam$^{-1}$} and appear in every single observation of a given region. The choice of \mbox{500 mJy beam$^{-1}$} is based on a desire to reach a relative brightness calibration of \mbox{$\sim$2\%} and the typical noise is \mbox{$\sim$10 mJy beam$^{-1}$} in each image (see Table \ref{regionlist}), yielding a SNR for the minimum brightness peaks of $\sim$50:1. We first calculate the average peak brightness over all epochs to remove the uncertainty in flux related to individual measurements and then normalise the observed source peak brightnesses in each observation to their respective averages. 

\vspace{3mm}

\item We then compare the brightness of each extracted source with respect to each of the others by taking the ratio of their normalised peak brightnesses and plotting the result for each epoch (for example, see Figure \ref{Rfig}). If a given pair of sources gets brighter or dimmer together from epoch to epoch, due to calibration uncertainties, we would expect the ratio of their normalised peak brightnesses over time to show little scatter.  

\vspace{3mm}

\item We next measure the amount of scatter in the ratio between two sources by calculating the standard deviation in the ratio of normalised peak brightnesses over time. For every pair of sources, we can plot the measured standard deviation. In Figure \ref{stdevplot}, we show the results of a simple model of the standard deviations measured for pairs of sources by applying a Gaussian error of the value indicated to 1000 sources of peak brightness 1.0 and comparing their expected normalised peak brightness ratios over eight epochs. Overlaid on this figure are the measured standard deviations of the normalised source peak brightness ratios for all nine potential calibrator sources in all eight observations of the Ophiuchus Core region. Note that as more epochs are observed, the central part of the curve (where the majority of the data points lie) regresses to a value of $\sqrt{2}\times\mathrm{\:error}$.

\vspace{3mm}

\item We next identify the largest set of sources wherein every pair has a standard deviation below a threshold set to 0.06. We call this set of stable sources a \textit{Family}. We choose 0.06 through comparison with model curves in Figure \ref{stdevplot}, and as a compromise between decreasing the number of family members versus increasing the reliability of the calibration. Thus, all \textit{Family} members satisfy the threshold when compared against each other. These sources are considered non-varying and appropriate for the relative flux calibration of each epoch. This threshold was chosen after extensive testing for the optimal number of sources contained within the \textit{Family} across all observed regions. 

\vspace{3mm}

\item For each epoch, the flux calibration factor by which we divide every pixel is the average normalised peak brightness over all the sources within the \textit{Family} during that observation. These factors are plotted in the left panel of Figure \ref{fcf_fig}.  
In the figure, black indicates all observations taken before March 1$^\text{st}$, 2017, while grey indicates observations taken after the filters were changed on SCUBA-2 in November, 2016. The standard deviation of a Gaussian fit to all the data is $8\%$, as expected \citep{dempsey2013}.  

\vspace{3mm}

\item The uncertainty in the derived flux calibration factor is taken to be the error in a given measurement of the normalised peak brightness of an individual source which is calculated by finding the standard deviation of the normalised peak brightnesses of all the calibrator sources. This uncertainty is plotted in the right panel of Figure \ref{fcf_fig}. Note that the uncertainty peaks at approximately 2\%. This is, however, the uncertainty per source while the error in the mean scales with the square root of the number of calibrator sources detected. Again, all four tested reductions show consistent results (e.g. compare the \textit{R3} flux calibration results with \textit{R4} in Appendix \ref{obsappendix}), though the \textit{R3} reduction is most robust for the JCMT Transient Survey science goals.

\end{enumerate}

The highest flux calibration factor uncertainty is found in the NGC2024 region for the same reasons the residual offset in the aligned maps is larger; the calibrator sources identified in NGC2024 are found within the extended structure where our Gaussian fitting routine encounters more uncertainty. 
Overall, 
the flux calibration uncertainties are very low ($\sim2-3\%$), allowing the JCMT Transient Survey team to robustly detect variations in peak brightness of the most prominent sources to the level of $\sim10\%$. Previously, \cite{haubois2012} achieved this relative calibration accuracy over a 5 day monitoring campaign of Sagittarius A* using the Atacama Pathfinder Experiment instrument at the Llano de Chajnantor Observatory. These small uncertantites, however, are unprecedented in ground-based, single dish submillimetre observations for such a wide range of consistent observations. In addition, we fit many sources in each field which allows us to detect sufficiently significant brightness variations on the first epoch they occur. Note that since we determine each epoch's flux calibration factor in a relative sense, the SCUBA-2 filter change in 2016 does not affect our results. The absolute brightness measurements, however, depend on the flux conversion factors (see Section \ref{drsec}). The filter change is expected to cause a small but detectable change in these values, but this has not yet been quantified by the observatory.

Our calibration is only expected to improve as we include future epochs and refine our methods throughout the duration of the survey. With additional epochs, we will be able to co-add subsets of our observations to increase our sensitivity at the cost of a lower cadence. Presently, we are working to automate the flux calibration procedure such that it can also be run directly after an observation undergoes the data reduction and alignment procedures at the EAO. For a given observation, the aligned and flux calibrated data are presently available to team members within 24-48 hours.

\section{Discussion}
\label{discusssec}

\begin{figure*} 	
\centering
\subfloat{\label{}\includegraphics[width=9cm,height=7.8cm]{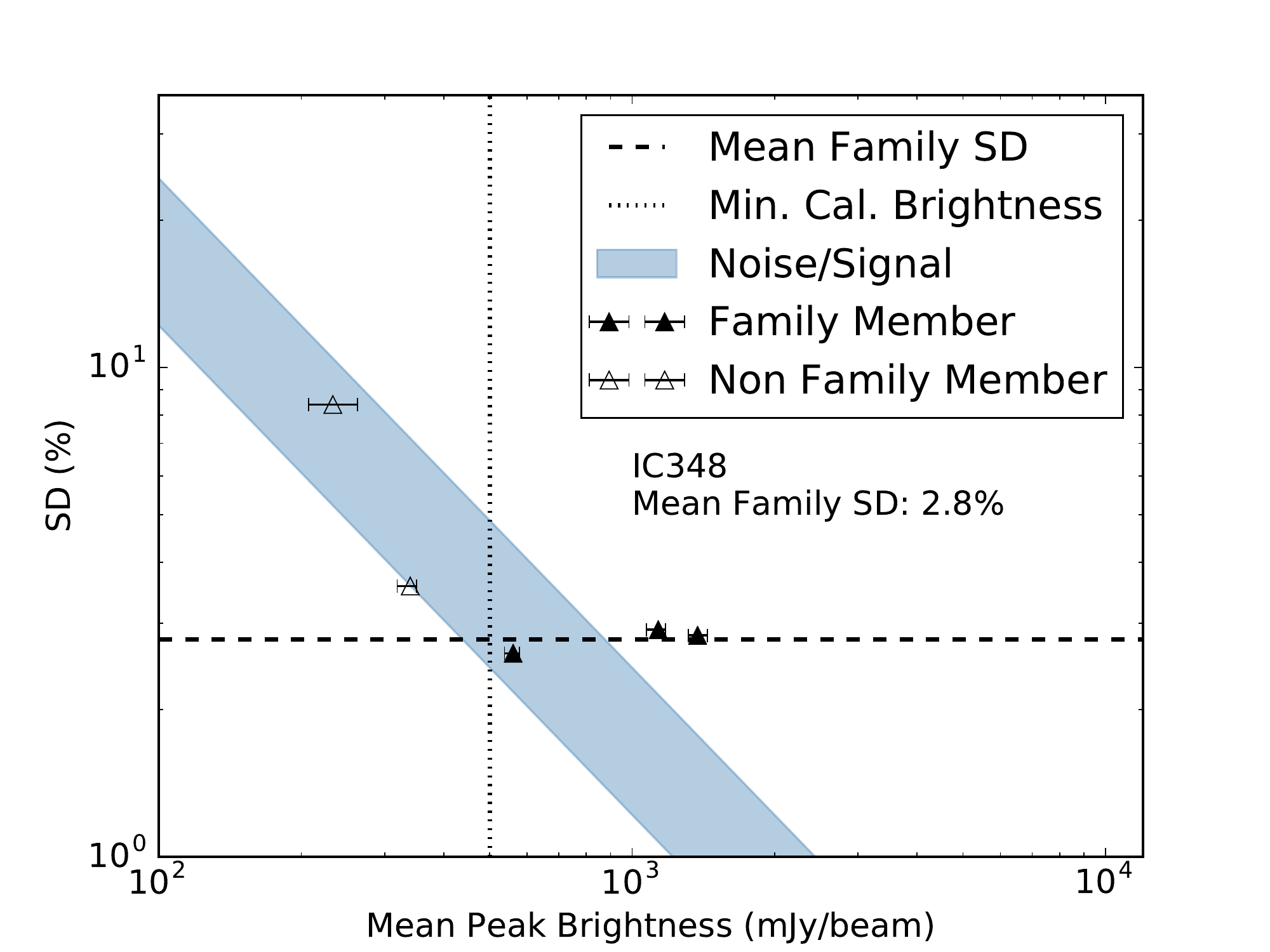}}
\subfloat{\label{}\includegraphics[width=9cm,height=7.8cm]{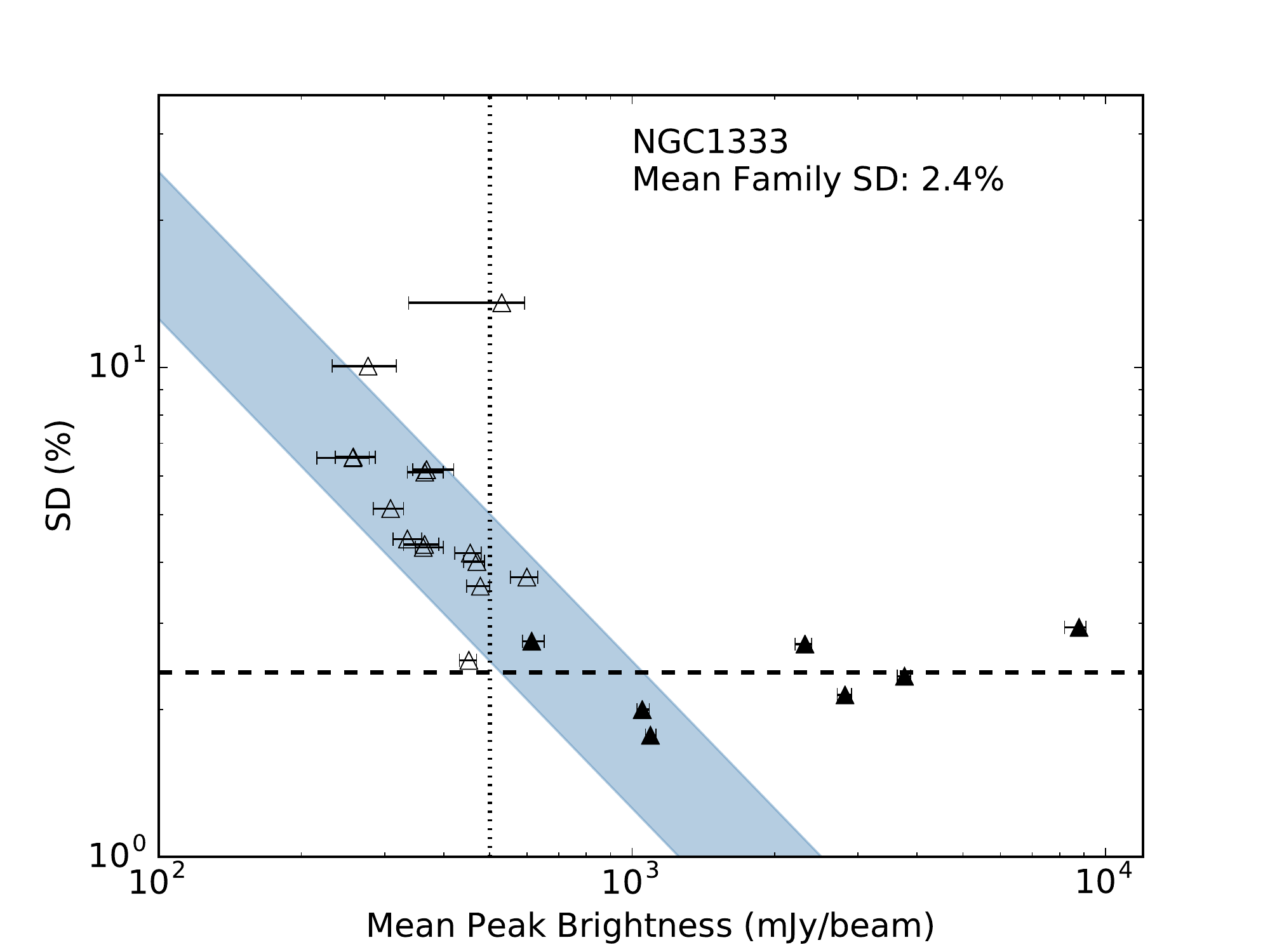}}\\
\subfloat{\label{}\includegraphics[width=9cm,height=7.8cm]{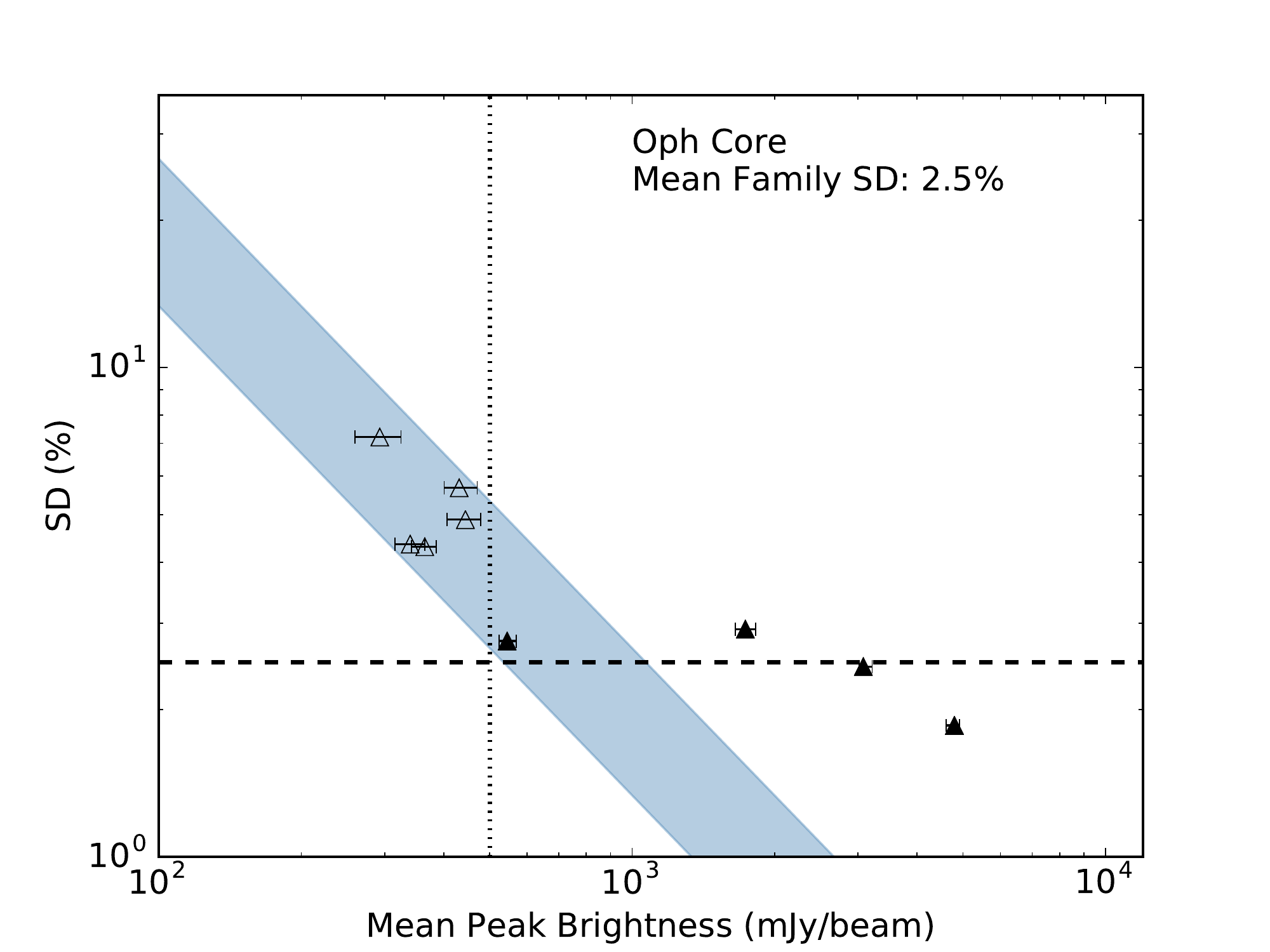}}
\subfloat{\label{}\includegraphics[width=9cm,height=7.8cm]{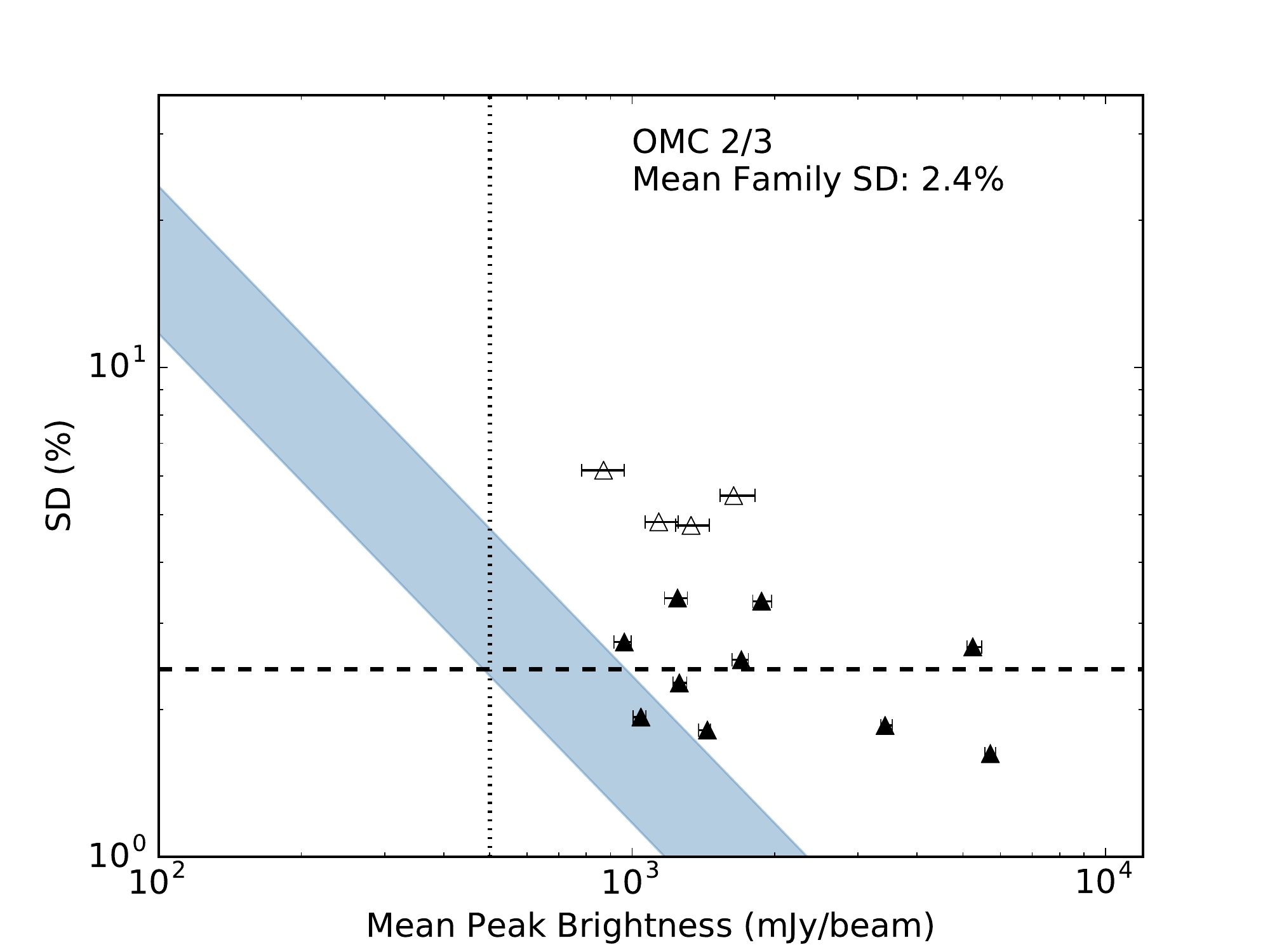}}
\caption{The standard deviation in the peak brightness versus the mean peak brightness of a source for four of the Transient fields. The horizontal errorbars indicate the range of peak brightnesses observed across all dates. Filled triangles represent \textit{Family} members while empty triangles represent other sources not included in the flux calibration. The vertical dotted line indicates the minimum brightness threshold to be considered a member of a \textit{Family}. The horizontal dashed line shows the average standard deviation in the mean peak brightness of all the \textit{Family} sources. The lower bound of the shaded region shows the average noise as a percentage of source peak brightness and the upper bound of the shaded region assumes the noise is higher by a factor of two.}
\label{SDvspeakbright1}
\end{figure*}

\begin{figure*} 	
\centering
\subfloat{\label{}\includegraphics[width=9cm,height=7.8cm]{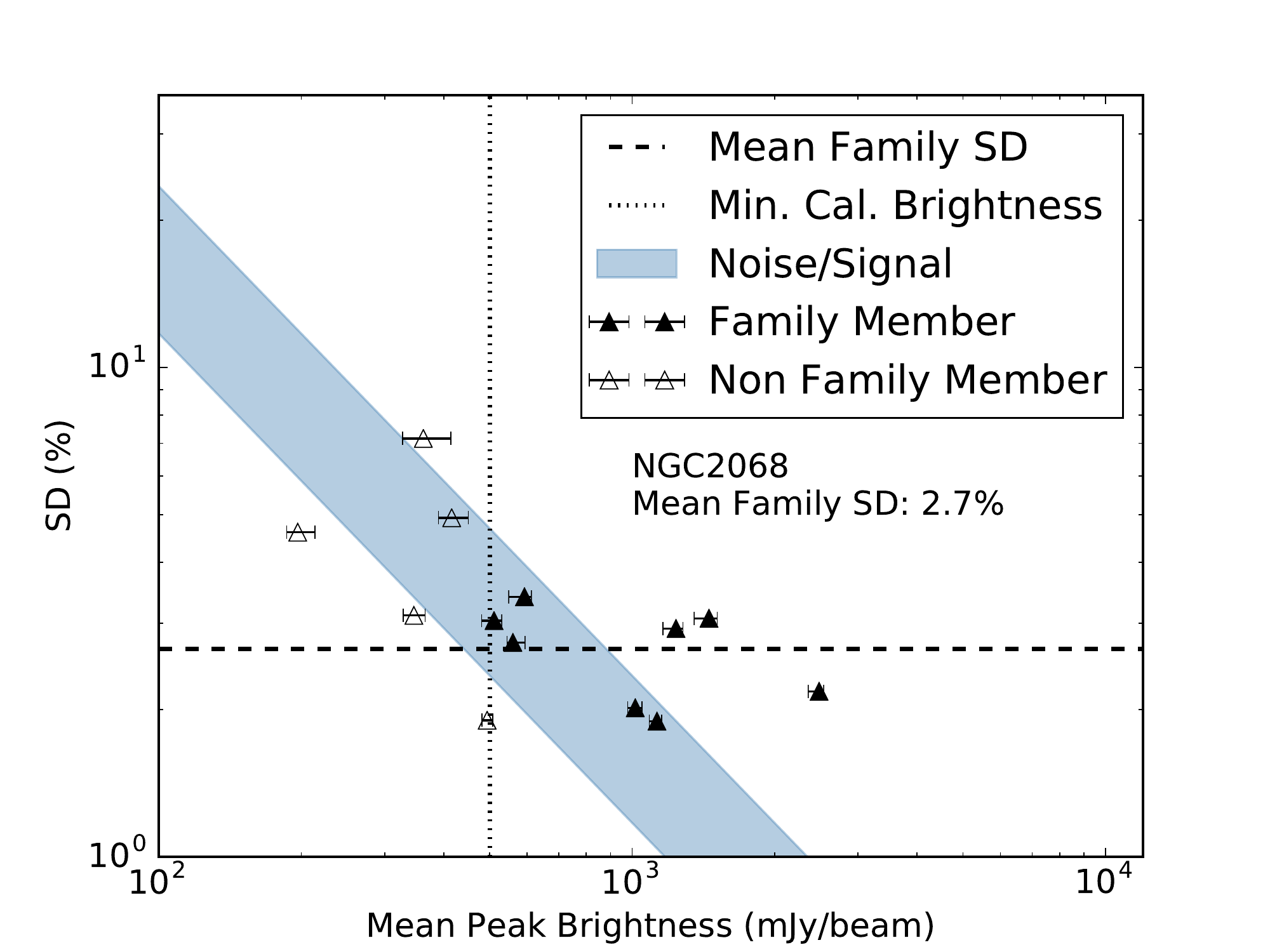}}
\subfloat{\label{}\includegraphics[width=9cm,height=7.8cm]{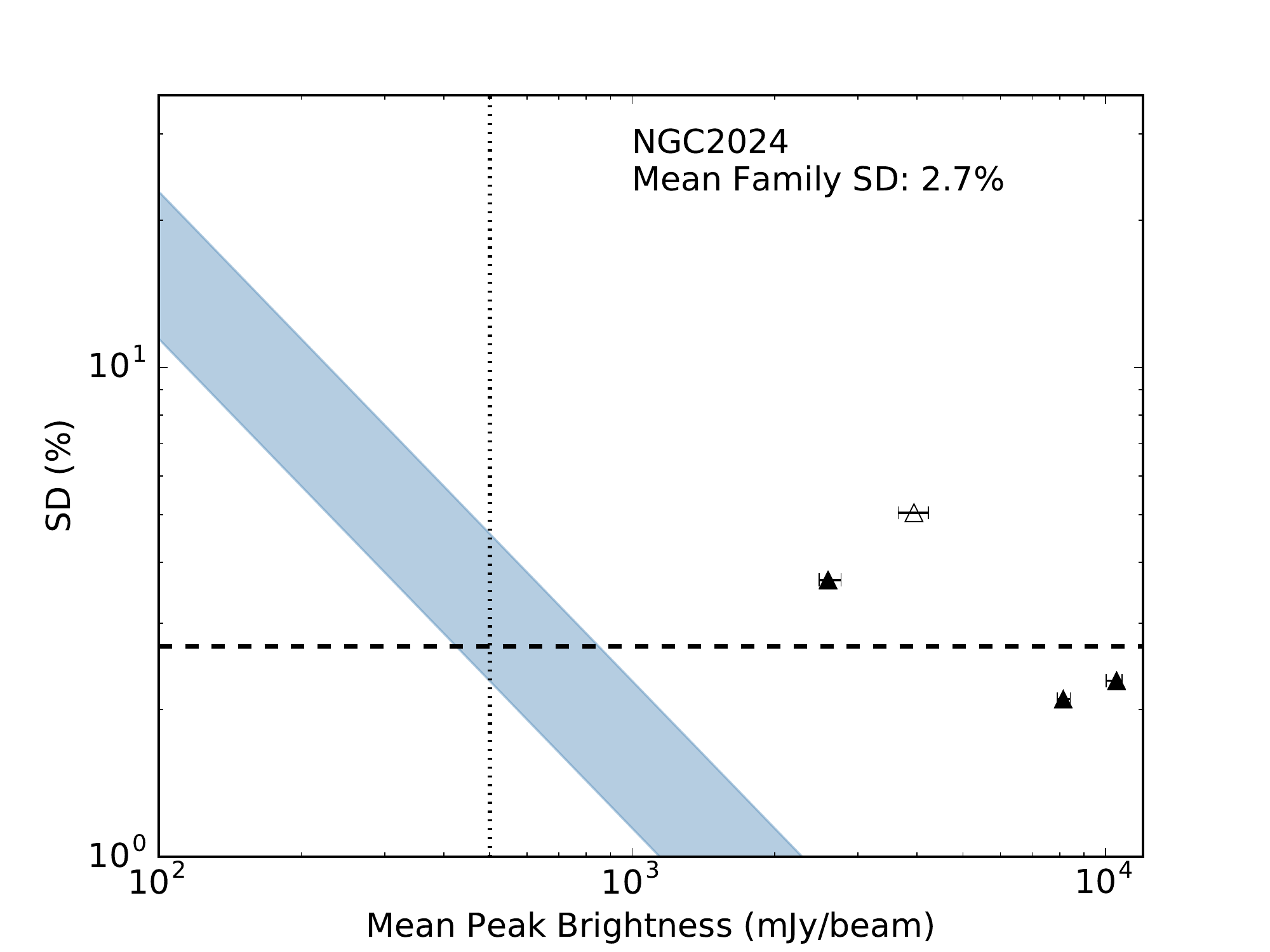}}\\
\subfloat{\label{}\includegraphics[width=9cm,height=7.8cm]{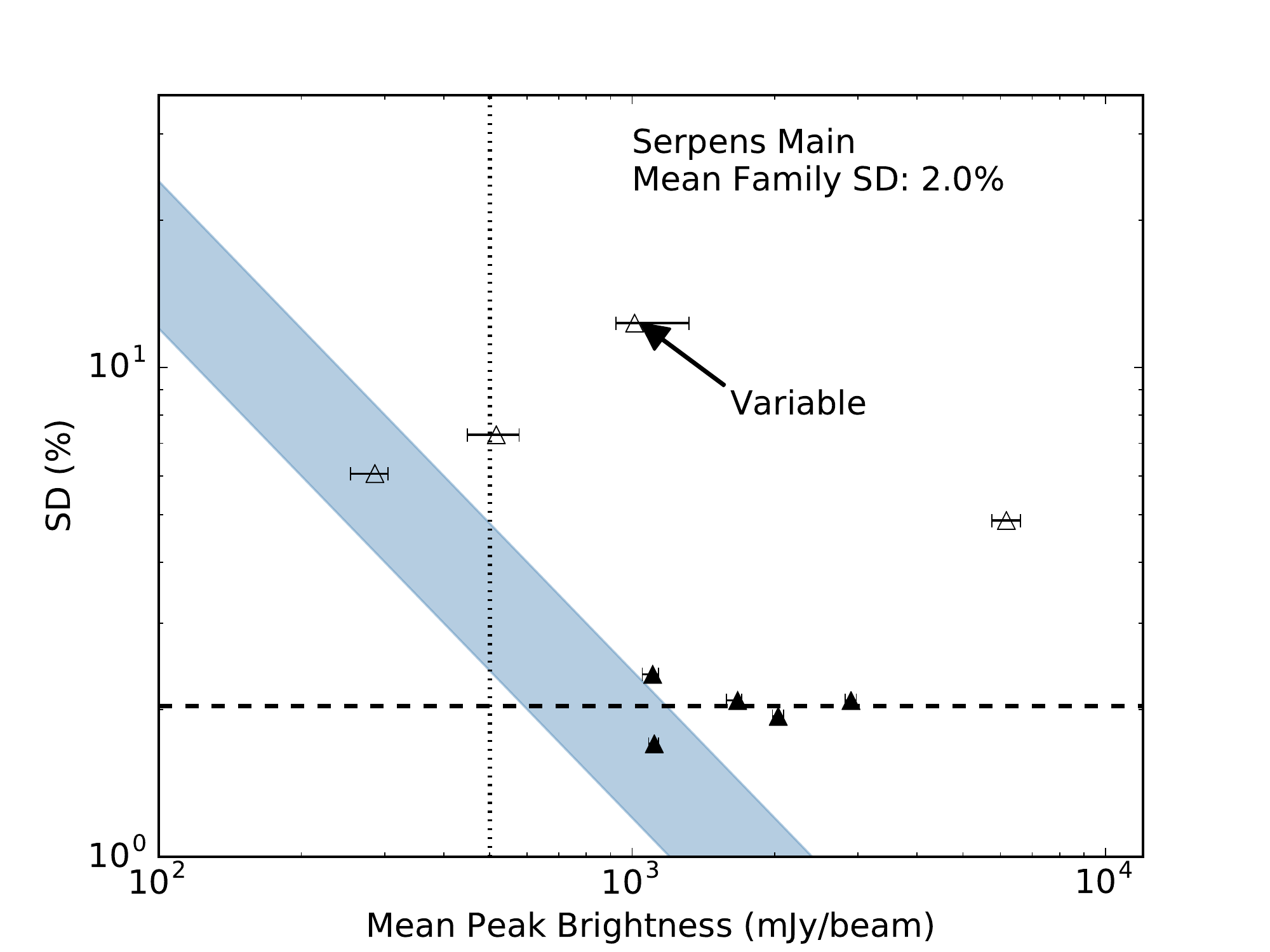}}
\subfloat{\label{}\includegraphics[width=9cm,height=7.8cm]{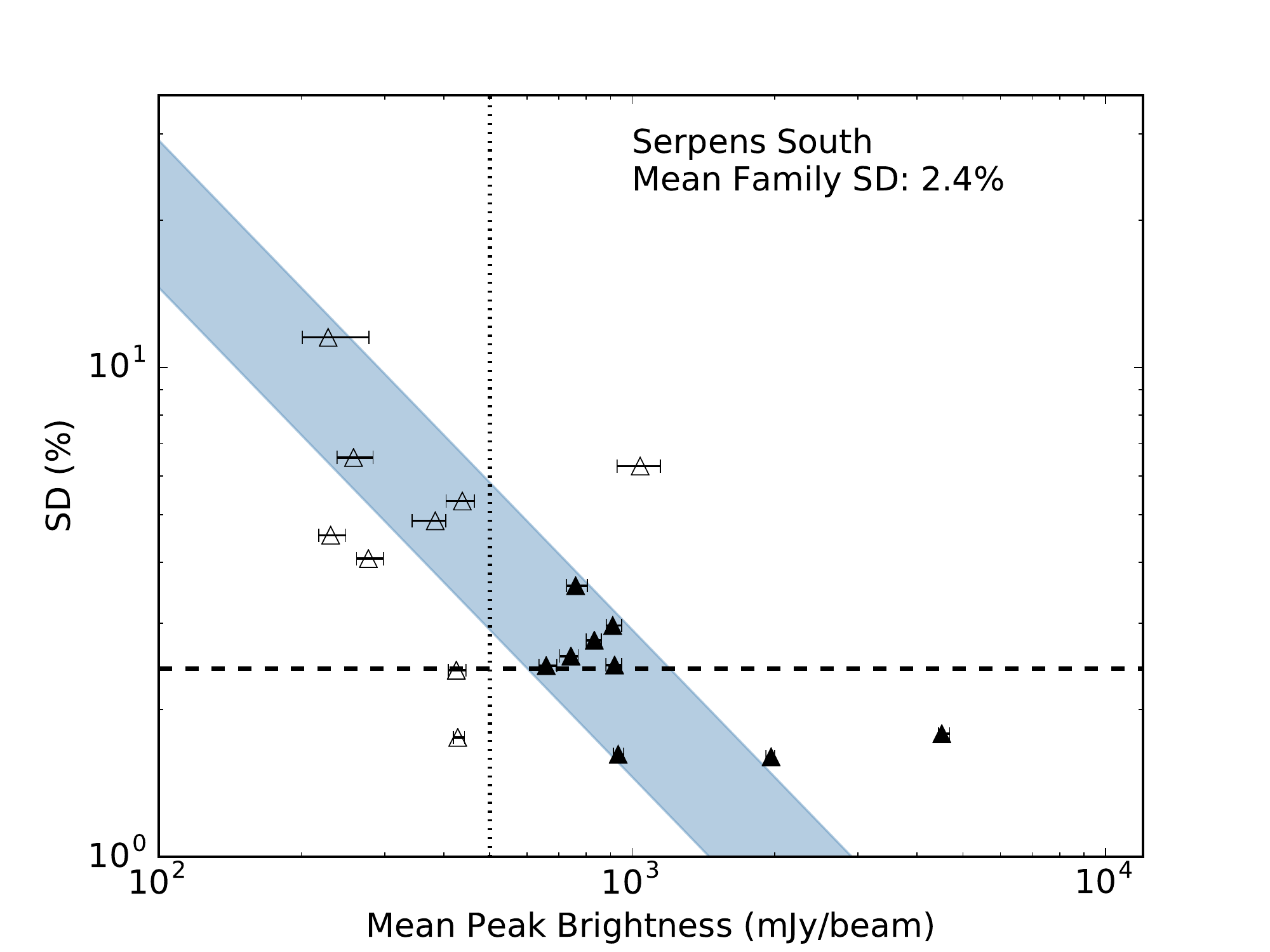}}
\caption{Same as Figure \ref{SDvspeakbright1}, showing the other four Transient fields.}
\label{SDvspeakbright2}
\end{figure*}

The goal of this calibration work is to extract robust, non-varying sources from SCUBA-2 maps and apply the spatial alignment and flux calibration methods. 
The majority of this process involves excluding sources from flux calibrator \textit{Families} which do not meet our set of criteria. These excluded sources, however, are of particular interest to the JCMT Transient Survey as 
they may be transient.

Figures \ref{SDvspeakbright1} and \ref{SDvspeakbright2} show the measured fractional variation in the fluxes (standard deviation) of each observed source across all epochs that it was observed, ordered by source brightness.
For a source to be included on these plots, it must be detected in every epoch of the given region. Thus, there are many additional, potentially interesting, submillimetre sources which are not included in these figures as our focus is on potential calibrators. A source may not be detected in a given epoch for two reasons: 1) It has properties near enough to the detection threshold that it is too faint or too extended on some observation dates but not on others (a source in a clustered environment is difficult to fit which may cause the shape to change). 2) The peak brightness of the source may vary such that it was too faint to be detected at the time of the reference observation but it was bright enough at a later date (or vice versa). The sources in both scenarios are of little importance to the flux calibration since we only want to use the most robust sources available. Detecting the sources in the second scenario is one of the goals of the JCMT Transient Survey and follow-up studies are currently underway to quantify their number and amplitude (for example, Mairs et al. in preparation). Including all observations of all eight regions, using our source extraction methods based on detecting compact structure and fitting Gaussian profiles with {\sc{Gaussclumps}} as detailed in Appendix \ref{gcappendix}, we see a total of 265 unique areas of significant, localised emission (see Table \ref{bigobstable}). This number is expected to vary depending on the source identification procedure used and the amount of data received. 

The lower bound of the shaded regions in Figures \ref{SDvspeakbright1} and \ref{SDvspeakbright2} show the average noise in each region (see Table \ref{regionlist}) as a percentage of the mean peak brightness while the upper bound represents the noise multiplied by a factor of two to take into consideration additional uncertainties due to, for example, the source fitting procedure. In general, we expect fainter sources to approximately follow the shaded region whereas we expect \textit{Family} members to lie further to the right and display low standard deviations dominated by the Gaussian fitting uncertainties. The vertical dotted line shows the minimum peak brightness threshold for a source to be considered a \textit{Family} member (\mbox{500 mJy beam$^{-1}$}) and the horizontal dashed line shows the mean standard deviation of the \textit{Family} members in that specific region.  

\begin{figure*} 	
\centering
\includegraphics[width=10cm,height=9cm]{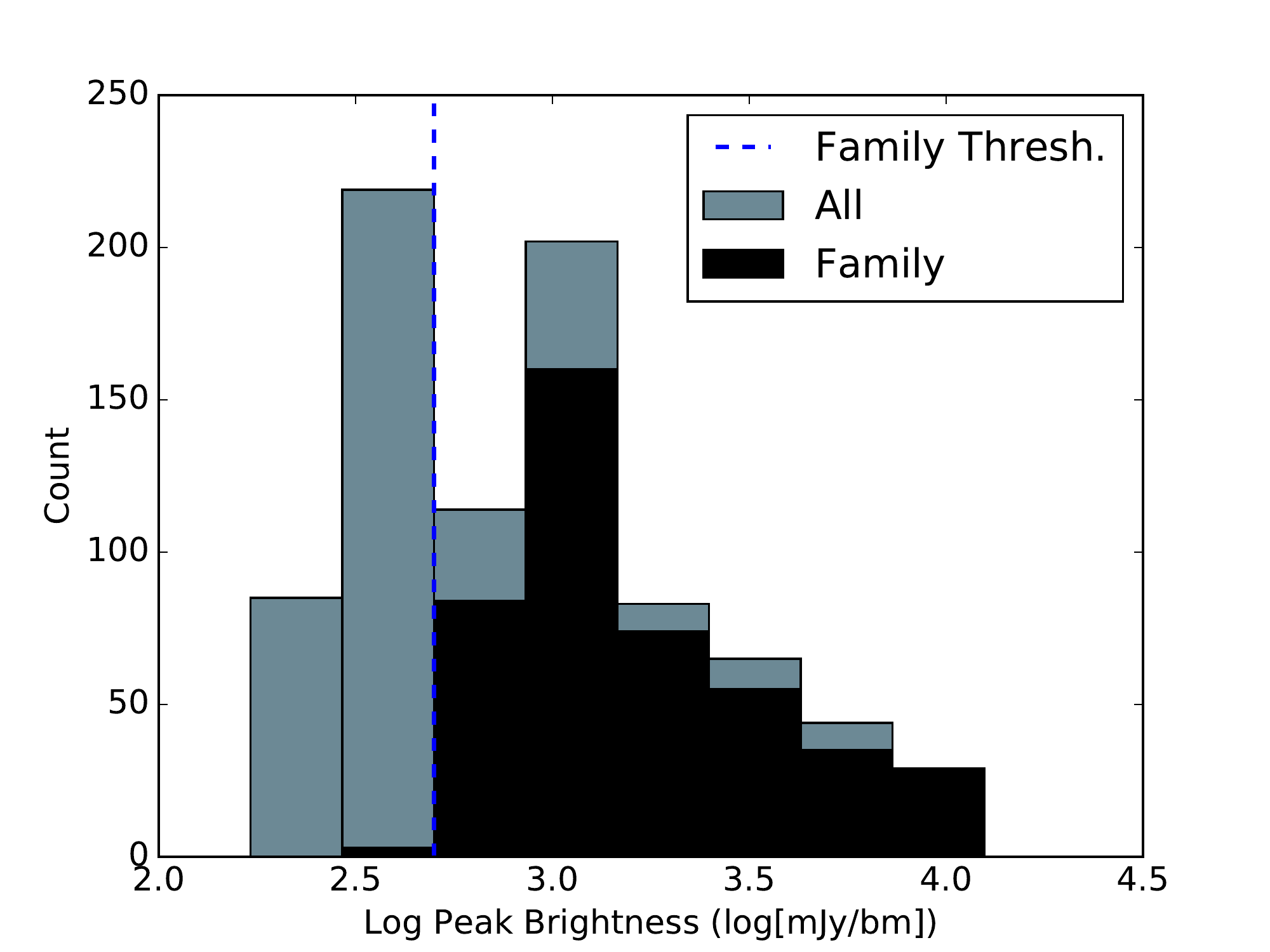}
\caption{Peak Brightnesses of all sources detected in every observation (grey) and only those included in the flux calibrator \textit{Families} (black). The dashed line indicates the minimum average brightness threshold required to be considered a \textit{Family} member. In some individual observations, \textit{Family} members have peak brightnesses which are slightly less than the threshold. Sources from every epoch of all eight fields prior to March 1$^\text{st}$, 2017 are included.} 
\label{lowpeak-lowconc}
\end{figure*}

Most of the sources behave as expected for objects which do not vary with time. There are, however, a few notable exceptions. The OMC 2/3 and NGC2024 fields are where localised Gaussian profiles are extracted from particularly clustered and confused emission (see, for example, Figure \ref{ngc2024cutout}). These two regions have the highest number of relatively bright sources not included in their \textit{Families}, most likely due to the source extraction procedure but also possibly due to intrinsic variability. {\sc{Gaussclumps}} is able to extract and fit well isolated, compact emission sources while sources extracted from clustered regions have more uncertainty. Depending on the morphology of the surrounding background structure, emission from multiple sources can be blended which causes some sources to deviate from Gaussian profiles, fluctuating in elongation from epoch to epoch as the algorithm attempts to separate the significant structure from the background. Examples of these sources include the two which meet the minimum brightness threshold but fail to be included in a \textit{Family} in NGC1333 (Figure \ref{SDvspeakbright1}, top right), 
the brightest source in the Serpens Main region (Figure \ref{SDvspeakbright2}, bottom left), and the source on the \textit{Family} brightness threshold in the Serpens Main region (Figure \ref{SDvspeakbright2}, bottom left). Since this paper is concerned with calibration, we simply ignore these more complicated sources. In future papers, however, we will adapt techniques to better identify variability in the most crowded regions in our fields. In general, sources which fail the flux calibrator criteria are lower peak brightness, as expected (see Figure \ref{lowpeak-lowconc}).

Another reason a source may significantly deviate from what we expect and fail to be included in a \textit{Family}  is that it is undergoing an observable, physical variation. Our relative flux calibration algorithm has been designed to exclude sources which are varying so that their contribution would not suppress the signal we seek to study. One variable source, 
has been identified (see Figure \ref{SDvspeakbright2}, bottom left) and verified over multiple observations in our dataset (for more detail on this source, refer to Yoo et al. in preparation). A careful but cursory analysis of each source that was detected in every observation and was excluded from a \textit{Family} has been carried out and no other clear and robust detections of significant variability have so far been identified. Investigations will continue, however, to uncover any long term trends. In addition, there are many  sources present in each map which are not presented in this paper. Analyses employing different source extraction methods as well as procedures which consider the variability of faint sources are currently underway (for example, Yoo et al. in preparation, Mairs et al. in preparation).  

\section{Conclusion}
\label{conclusionsec}

The primary goal of the JCMT Transient Survey is to detect variability in the brightness of deeply embedded protostars. The pointing accuracy of the JCMT is nominally 2 to 6 arcseconds while the nominal flux calibration uncertainty of \mbox{850 $\mu$m} SCUBA-2 data is $8\%$ (see Figures \ref{align_fig} and \ref{fcf_fig}). In order to dramatically increase our sensitivity to variable signals, we have developed a calibration pipeline which further spatially aligns multiple observations of a given field and provides a relative flux calibration correction for bright, compact sources. We use the algorithm {\sc{Gaussclumps}} (see Section \ref{postredcal} and Appendix \ref{gcappendix}) to extract locations and peak brightnesses of emission objects in the \mbox{850 $\mu$m} SCUBA-2 maps and we apply minimum brightness ($200\mathrm{\:mJy\:beam}^{-1}$ for the spatial alignment and $500\mathrm{\:mJy\:beam}^{-1}$ for the relative flux calibration) and maximum radius ($10\arcsec$) thresholds to ensure we have the best fit objects in our sample. These methods could be applied to any submillimetre data with multiple observations of bright, compact objects. Our main results can be summarised as follows:

\begin{enumerate}[labelindent=0pt,labelwidth=\widthof{\ref{last-item}},label=\arabic*.,itemindent=1em,leftmargin=!]

\item We thoroughly tested four different data reduction methods and found the most robust parameters for our scientific goals (Reduction \textit{R3}, see Section \ref{drsec}).

\vspace{3mm}

\item We achieve a sub-pixel alignment uncertainty of $\sim1\arcsec$ (see Figure \ref{align_fig} and Section \ref{imagealign}), improving on the pointing error of the telescope by a factor of $\sim4$. 

\vspace{3mm}

\item We achieve a relative flux calibration factor uncertainty of $2-3\%$ for bright sources (see Figure \ref{align_fig} and Section \ref{imagealign}), improving on the native, absolute flux calibration uncertainty by a factor of $\sim3$. This is unprecedented in ground-based, single dish submillimetre observations. 

\vspace{3mm}

\item By analysing the bright sources that are not included in flux calibrator \textit{Families}, we have noted a variable source at \mbox{850 $\mu$m} (Yoo et al. in preparation) and identified good source extraction practices for further analysis  to improve the calibration procedure.

\end{enumerate}

The JCMT Transient Survey is expected to continue through at least January, 2019, increasing the number of observed epochs for each region by a factor of three to about thirty. Throughout this time, we will be working to improve the data reduction and calibration procedures (see Appendix \ref{kevinappendix}) in order to detect fainter signals and working to achieve similar results for the relative flux calibration uncertainty at \mbox{450 $\mu$m}. By the end of the survey, we will have the deepest submillimetre maps of these eight regions which will create many opportunities for additional science, including co-adding across several epochs to uncover variability in fainter sources, but with a lower cadence. 

\section*{Acknowledgements}

Steve Mairs was partially supported by the Natural Sciences and
Engineering Research Council (NSERC) of Canada graduate scholarship
program. Doug Johnstone is supported by the National Research Council
of Canada and by an NSERC Discovery Grant. 

\vspace{3mm}

The authors wish to recognise and acknowledge the very significant cultural role and
reverence that the summit of Maunakea has always had within the indigenous Hawaiian
community. We are most fortunate to have the opportunity to conduct observations from
this mountain. 
The James Clerk Maxwell Telescope
is now operated by the East Asian Observatory on behalf of
The National Astronomical Observatory of Japan, Academia Sinica
Institute of Astronomy and Astrophysics, the Korea Astronomy
and Space Science Institute, the National Astronomical Observatories
of China and the Chinese Academy of Sciences (grant no.
XDB09000000), with additional funding support from the Science
and Technology Facilities Council of the United Kingdom and participating
universities in the United Kingdom and Canada. The identification number for the JCMT Transient Survey under which the SCUBA-2
data used in this paper is M16AL001. 
The authors thank the JCMT staff for their support
of the data collection and reduction efforts. The {\sc{starlink}} software is supported by the East Asian Observatory. This research has made use of NASA's Astrophysics Data System and the facilities of the Canadian Astronomy Data Centre operated by the National Research Council of Canada with the support of the Canadian Space Agency. This research used the services
of the Canadian Advanced Network for Astronomy Research (CANFAR, specifically VOSpace) which in turn is
supported by CANARIE, Compute Canada, University of Victoria, the National Research
Council of Canada, and the Canadian Space Agency. This research made use of {\sc{APLpy}}, an open-source plotting package for Python hosted at http://aplpy.github.com, and {\sc{matplotlib}}, a 2D plotting library for Python \citep{matplotlib}. The authors would like to thank the anonymous referee for providing useful comments that strengthened this work. 

\facility{JCMT (SCUBA-2) \citep{holland2013}}

\software{Starlink \citep{currie2014}, Astropy \citep{astropy}, Python version 3.5, APLpy \citep{aplpy}, Matplotlib \citep{matplotlib}}

\appendix


\section{Observational Data}
\label{obsappendix}

\begin{figure*} 	
\centering
\subfloat{\label{}\includegraphics[width=9cm,height=7.8cm]{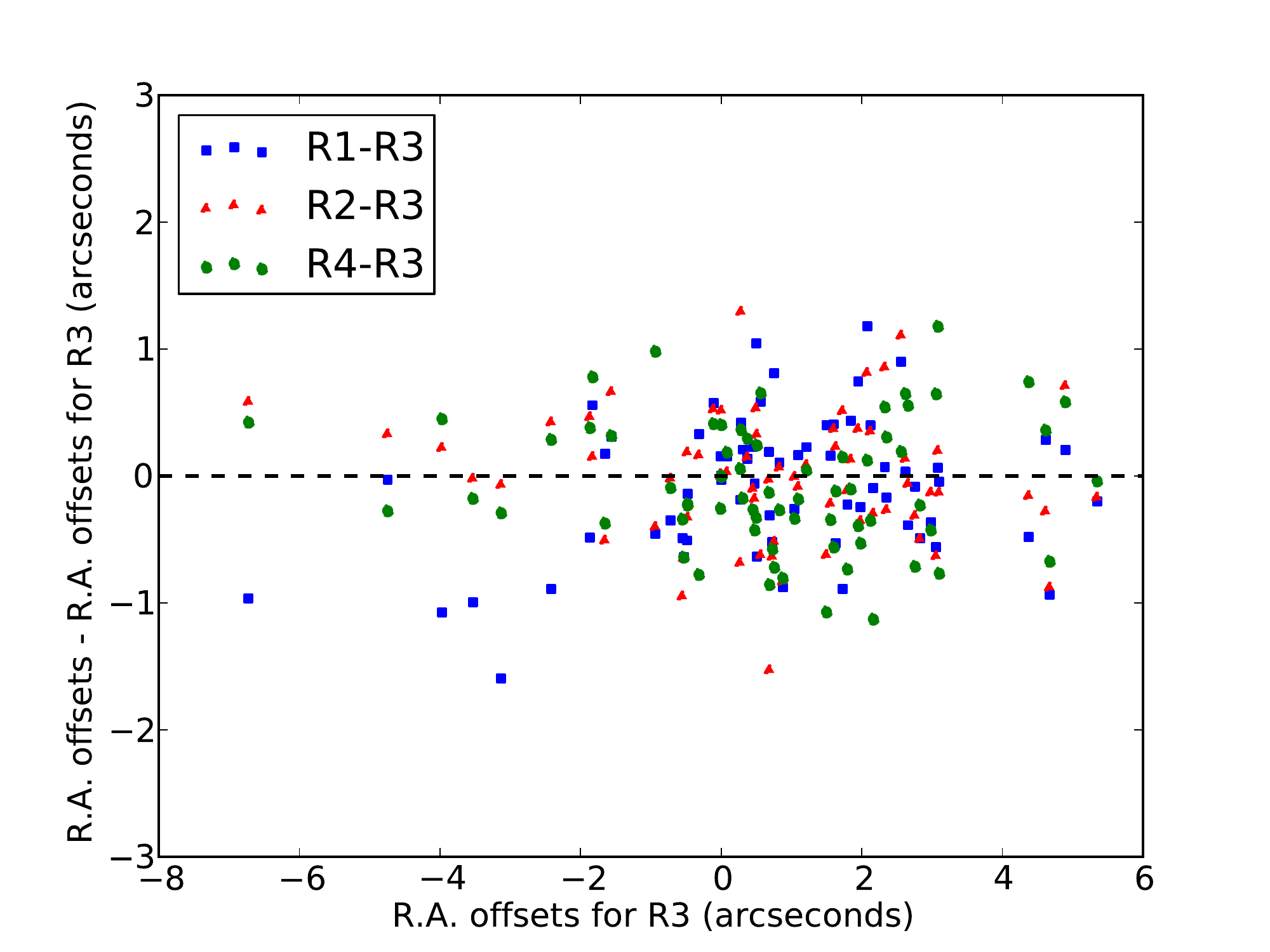}}
\subfloat{\label{}\includegraphics[width=9cm,height=7.8cm]{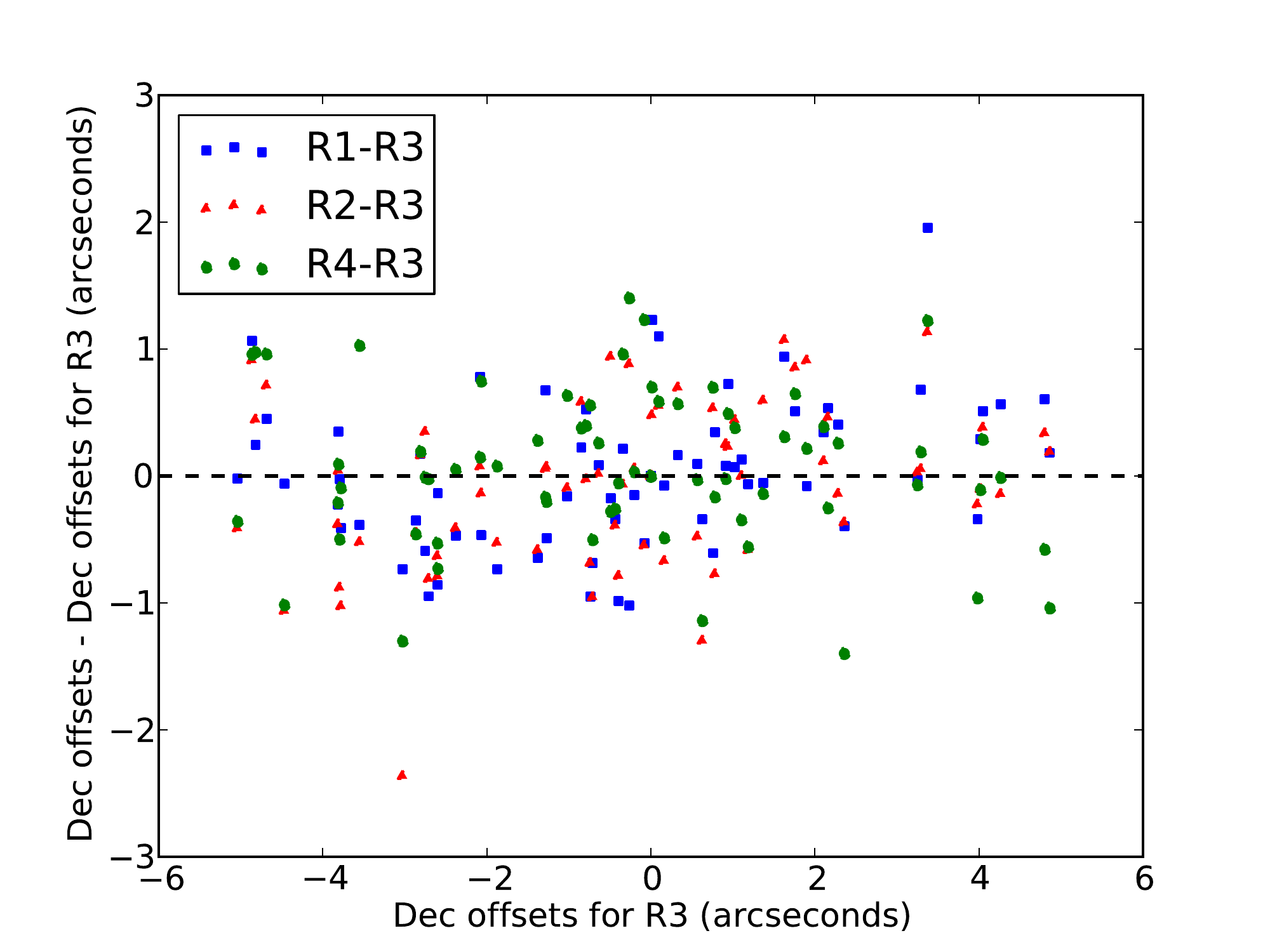}}
\caption{\textit{Left:} The derived right ascension offsets measured in each of the four reductions compared to reduction \textit{R3}. \textit{Right:} Same as left, but showing the declination offsets. In both panels the x-axis is used  to discriminate between observations and to show that our ability to align is independent of the original pointing error at the telescope.}
\label{align-app}
\end{figure*}

\begin{figure*} 	
\centering
\includegraphics[width=12cm,height=10cm]{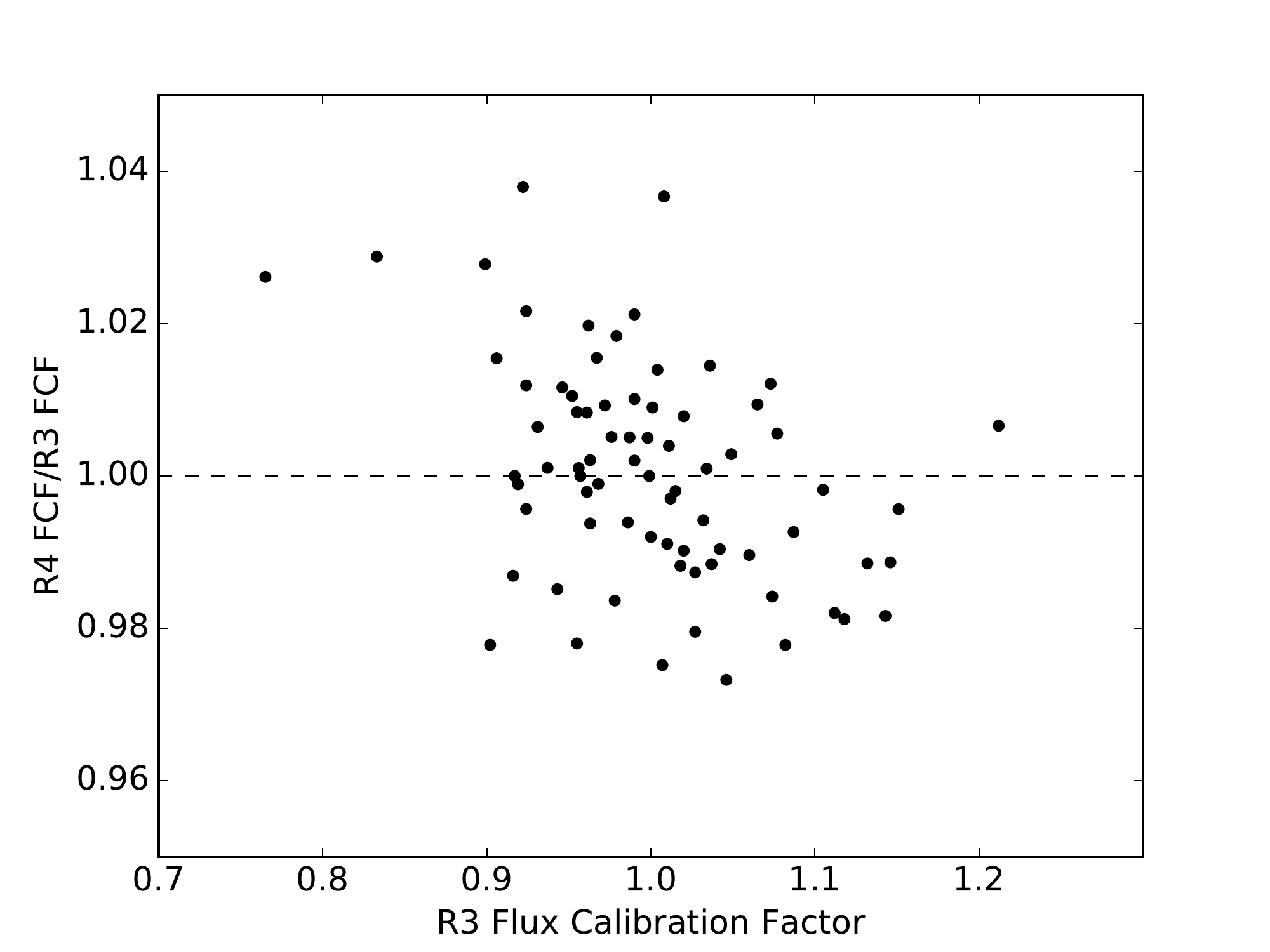}
\caption{The derived flux calibration factors for compact emission sources for all observations of the eight regions using the \textit{R4} reduction normalised by the flux calibration factors derived using the \textit{R3} reduction. As in Figure \ref{align-app}, the x-axis is used  to discriminate between observations.}
\label{FCF-app}
\end{figure*}

Table \ref{bigobstable} presents detailed information about all of the JCMT Transient Survey observations taken between December 22$^\text{nd}$, 2015 (the beginning of the survey) and March 1$^\text{st}$, 2017. Reduction \textit{R3} is used in each case. Figure \ref{align-app} highlights that the four tested data reduction methods produce consistent spatial alignment results overall while Figure \ref{FCF-app} shows that the \textit{R3} and \textit{R4} (and, by extension, \textit{R1} and \textit{R2}) relative flux calibration results are also consistent. 





\begin{centering}
\begin{longtable}{|c|c|c|c|c|c|c|c|}
\caption{A summary of all JCMT Transient Survey observations obtained between December 22$^\text{nd}$, 2015 (the beginning of the survey) and March 1$^\text{st}$, 2017.}
\label{bigobstable}\\
\hline
Region & Date & Scan & $\tau^{a}$ & \multicolumn{1}{|p{2.25cm}|}{\centering \mbox{850 $\mu$m} Noise$^{b,c}$\\ (mJy\: beam$^{-1}$)} &  \multicolumn{1}{|p{2.25cm}|}{\centering Number of Sources Above \\ 10$\sigma_{rms}$} & \multicolumn{1}{|p{2.25cm}|}{\centering Number of \\ Alignment Sources} & \multicolumn{1}{|p{2.25cm}|}{\centering Number of \\ Family Members} \\ \hline
\endfirsthead

\multicolumn{8}{c}
{{\bfseries \tablename\ \thetable{} -- continued from previous page}} \\ \hline
Region & Date & Scan & $\tau^{a}$ & \multicolumn{1}{|p{2.25cm}|}{\centering \mbox{850 $\mu$m} Noise$^{b,c}$\\ (mJy\: beam$^{-1}$)} &  \multicolumn{1}{|p{2.25cm}|}{\centering Number of Sources Above \\ 10$\sigma_{rms}$} & \multicolumn{1}{|p{2.25cm}|}{\centering Number of \\ Alignment Sources} & \multicolumn{1}{|p{2.25cm}|}{\centering Number of \\ Family Members} \\ \hline 
\endhead

\hline \multicolumn{8}{|r|}{{Continued on next page}} \\ \hline
\endfoot

\hline \hline
\endlastfoot

\hline\hline
IC348 & 20151222 & 19 & 0.06 & 12.54 & 9 & 6 & 3\\
IC348 & 20160115 & 22 & 0.07 & 9.99 & 12 & 6 & 3\\
IC348 & 20160205 & 18 & 0.04 & 12.79 & 12 & 6 & 3\\
IC348 & 20160226 & 20 & 0.05 & 12.39 & 13 & 5 & 3\\
IC348 & 20160318 & 27 & 0.05 & 11.1 & 12 & 5 & 3\\
IC348 & 20160417 & 09 & 0.04 & 11.0 & 13 & 5 & 3\\
IC348 & 20160826 & 40 & 0.08 & 14.33 & 12 & 6 & 3\\
IC348 & 20161126 & 22 & 0.05 & 12.29 & 14 & 6 & 3\\
IC348 & 20170209 & 28 & 0.09 & 13.17 & 12 & 5 & 3\\
NGC1333 & 20151222 & 18 & 0.06 & 12.22 & 39 & 36 & 7\\
NGC1333 & 20160115 & 10 & 0.08 & 11.76 & 40 & 29 & 7\\
NGC1333 & 20160205 & 17 & 0.04 & 12.99 & 39 & 29 & 7\\
NGC1333 & 20160229 & 17 & 0.04 & 11.46 & 42 & 26 & 7\\
NGC1333 & 20160325 & 11 & 0.06 & 11.58 & 33 & 25 & 7\\
NGC1333 & 20160802 & 31 & 0.09 & 12.31 & 43 & 28 & 7\\
NGC1333 & 20160830 & 48 & 0.09 & 14.05 & 39 & 33 & 7\\
NGC1333 & 20161119 & 88 & 0.07 & 9.65 & 38 & 28 & 7\\
NGC1333 & 20161126 & 21 & 0.05 & 12.78 & 36 & 27 & 7\\
NGC1333 & 20170206 & 29 & 0.12 & 13.85 & 42 & 26 & 7\\
NGC2024 & 20151226 & 49 & 0.12 & 12.58 & 12 & 10 & 3\\
NGC2024 & 20160116 & 22 & 0.06 & 9.8 & 15 & 7 & 3\\
NGC2024 & 20160206 & 13 & 0.04 & 11.88 & 21 & 7 & 3\\
NGC2024 & 20160229 & 22 & 0.04 & 13.01 & 17 & 7 & 3\\
NGC2024 & 20160325 & 21 & 0.06 & 11.35 & 19 & 8 & 3\\
NGC2024 & 20160329 & 10 & 0.05 & 9.1 & 19 & 7 & 3\\
NGC2024 & 20160427 & 12 & 0.05 & 11.91 & 9 & 6 & 3\\
NGC2024 & 20160826 & 29 & 0.09 & 12.84 & 23 & 8 & 3\\
NGC2024 & 20161119 & 99 & 0.07 & 9.13 & 16 & 9 & 3\\
NGC2024 & 20161126 & 53 & 0.06 & 10.29 & 18 & 8 & 3\\
NGC2024 & 20170206 & 25 & 0.11 & 11.68 & 15 & 8 & 3\\
NGC2068 & 20151226 & 52 & 0.12 & 13.4 & 32 & 23 & 8\\
NGC2068 & 20160116 & 27 & 0.06 & 9.65 & 33 & 21 & 8\\
NGC2068 & 20160206 & 15 & 0.05 & 12.08 & 31 & 21 & 8\\
NGC2068 & 20160229 & 13 & 0.04 & 12.21 & 29 & 18 & 8\\
NGC2068 & 20160329 & 11 & 0.06 & 10.82 & 33 & 22 & 8\\
NGC2068 & 20160427 & 13 & 0.05 & 12.68 & 28 & 16 & 8\\
NGC2068 & 20160827 & 53 & 0.08 & 11.8 & 31 & 21 & 8\\
NGC2068 & 20161120 & 88 & 0.09 & 11.98 & 30 & 20 & 8\\
NGC2068 & 20161126 & 56 & 0.06 & 10.16 & 30 & 21 & 8\\
NGC2068 & 20170206 & 17 & 0.11 & 12.76 & 32 & 20 & 8\\
OMC 2/3 & 20151226 & 36 & 0.11 & 12.48 & 55 & 51 & 10\\
OMC 2/3 & 20160116 & 19 & 0.06 & 9.94 & 30 & 22 & 10\\
OMC 2/3 & 20160206 & 12 & 0.04 & 12.03 & 26 & 21 & 10\\
OMC 2/3 & 20160229 & 11 & 0.04 & 11.8 & 45 & 30 & 10\\
OMC 2/3 & 20160325 & 15 & 0.06 & 10.74 & 29 & 20 & 10\\
OMC 2/3 & 20160422 & 11 & 0.05 & 11.4 & 28 & 23 & 10\\
OMC 2/3 & 20160826 & 20 & 0.11 & 15.21 & 25 & 17 & 10\\
OMC 2/3 & 20161126 & 52 & 0.06 & 9.74 & 56 & 40 & 10\\
OMC 2/3 & 20170206 & 21 & 0.12 & 12.13 & 43 & 32 & 10\\
Oph Core & 20160115 & 84 & 0.07 & 11.9 & 26 & 23 & 4\\
Oph Core & 20160205 & 63 & 0.04 & 12.47 & 22 & 13 & 4\\
Oph Core & 20160226 & 51 & 0.05 & 11.08 & 27 & 17 & 4\\
Oph Core & 20160319 & 65 & 0.04 & 12.56 & 27 & 16 & 4\\
Oph Core & 20160417 & 43 & 0.04 & 12.03 & 24 & 16 & 4\\
Oph Core & 20160521 & 34 & 0.08 & 15.03 & 27 & 15 & 4\\
Oph Core & 20160826 & 11 & 0.11 & 17.56 & 24 & 14 & 4\\
Oph Core & 20170206 & 83 & 0.11 & 14.2 & 22 & 16 & 4\\
Serpens Main & 20160202 & 54 & 0.09 & 12.11 & 23 & 21 & 5\\
Serpens Main & 20160223 & 50 & 0.05 & 11.68 & 22 & 18 & 5\\
Serpens Main & 20160317 & 51 & 0.04 & 12.2 & 21 & 14 & 5\\
Serpens Main & 20160415 & 46 & 0.04 & 11.82 & 22 & 16 & 5\\
Serpens Main & 20160521 & 39 & 0.08 & 14.01 & 22 & 15 & 5\\
Serpens Main & 20160722 & 23 & 0.1 & 11.49 & 23 & 14 & 5\\
Serpens Main & 20160827 & 12 & 0.09 & 11.32 & 24 & 15 & 5\\
Serpens Main & 20160929 & 12 & 0.09 & 11.95 & 18 & 13 & 5\\
Serpens Main & 20170222 & 70 & 0.1 & 11.47 & 26 & 12 & 5\\
Serpens South & 20160202 & 58 & 0.09 & 11.27 & 39 & 35 & 9\\
Serpens South & 20160223 & 65 & 0.05 & 18.66 & 39 & 32 & 9\\
Serpens South & 20160317 & 52 & 0.04 & 11.41 & 34 & 25 & 9\\
Serpens South & 20160415 & 48 & 0.04 & 11.57 & 38 & 29 & 9\\
Serpens South & 20160521 & 44 & 0.07 & 12.61 & 38 & 27 & 9\\
Serpens South & 20160721 & 11 & 0.08 & 19.42 & 41 & 31 & 9\\
Serpens South & 20160827 & 17 & 0.09 & 17.05 & 43 & 32 & 9\\
Serpens South & 20160929 & 18 & 0.08 & 11.34 & 39 & 30 & 9\\
Serpens South & 20170222 & 81 & 0.1 & 17.68 & 36 & 28 & 9\\
\end{longtable}
\end{centering}

\begin{flushleft}
$^{a}$The average 225 GHz zenith opacity measured throughout the duration of the observation.\\ 
$^{b}$These measurements of the \mbox{850 $\mu$m} noise ($\sigma_{rms}$) levels are based on a point source detection in a single observation using 3$\arcsec$ pixels and a 14.6$\arcsec$ FWHM beam.\\
$^{c}$The reduction method R3 was used to derive these noise estimates (see Section \ref{drsec}).\\
\end{flushleft}


\section{Gaussclumps}
\label{gcappendix}

{\sc{Gaussclumps}}  \citep{stutzki1990} identifies first the brightest peak in a user-supplied map and fits a 2D Gaussian profile to the surrounding structure with a least-squares method. If the structure is deemed to be real (i.e., not a spurious detection of a noise spike, a test which is based on a series of user-defined parameters), the fit is subtracted from the data and the algorithm iteratively identifies and fits a Gaussian to the next brightest peak until all of the significant structure is identified. The algorithm is designed to weight the Gaussian fits to smaller-scale structures (at least the size of the instrument beam) such that locally peaked objects are preferred over underlying, larger-scale features. Identified sources are allowed to overlap. 

There are many user defined parameters that control how {\sc{Gaussclumps}} identifies structure as well as when it terminates after successfully extracting all of the emission found to be significant. Here, we summarise the main parameters we have used. For the parameters not listed, we simply use the default values:

\begin{enumerate}[labelindent=0pt,labelwidth=\widthof{\ref{last-item}},label=\arabic*.,itemindent=1em,leftmargin=!]

\item \mbox{\textit{BACKOFF} = True}. This parameter subtracts the background level from each identified emission source before calculating the sizes and centroid positions. 

\vspace{3mm}

\item \mbox{\textit{FwhmBeam} = 5}. This parameter defines the full width at half maximum (FWHM) size of the JCMT beam in pixels. No Gaussians which are smaller than this will be considered robust. 
For the \mbox{850 $\mu$m} data, we use a pixel size of 3$\arcsec$ while the data smoothed with a 6$\arcsec$ Gaussian kernel has a  FWHM beam size of 15.8$\arcsec$. 

\vspace{3mm}

\item \mbox{\textit{MaxBad} = 0}. This parameter determines the maximum fraction of bad pixels which can be included in an identified source.

\vspace{3mm}

\item \mbox{\textit{MaxNF} = 150}. This parameter determines the maximum number of times the chi-squared between the Gaussian model and the data will be iteratively evaluated and adjusted.

\vspace{3mm}

\item \mbox{\textit{MaxSkip} = 20}. This parameter sets the maximum number of consecutive failures to fit Gaussians. If {\sc{Gaussclumps}} fails to fit a Gaussian to the data more than 20 consecutive times, the algorithm terminates.

\vspace{3mm}

\item \mbox{\textit{Thresh} = 10.0}. This parameter defines the minimum peak brightness of a fitted Gaussian in units of the measured root mean square (RMS) noise. Note that we also measure the RMS noise for each map individually and supply that value to {\sc{Gaussclumps}}' \textit{RMS} parameter. 

\end{enumerate}

The typical RMS noise in a given \mbox{850 $\mu$m} observation is \mbox{$\sim10\mathrm{\:mJy\: beam}^{-1}$} (see Tables \ref{regionlist} and \ref{bigobstable}), so we catalogue sources with peak brightnesses above \mbox{$\sim100\mathrm{\:mJy\: beam}^{-1}$}. 
Once we obtain the results from {\sc{Gaussclumps}}, we apply an additional cull to the catalogue to select the brightest, locally peaked objects  to select the image alignment and flux calibrator sources. Based on a series of tests varying the minimum brightness threshold and maximum source radius of identified objects, we select sources with peak brightnesses greater than \mbox{$200$ mJy\: beam$^{-1}$} and radii less than $10\arcsec$. We define the radius of a source to be \mbox{$r = \sqrt{\mathrm{FWHM_{1}}\times \mathrm{FWHM}_{2}}/2$} where the FWHM$_{N}$ terms are the full widths at half maximum of the two dimensional Gaussian. For the flux calibrator sources, we select sources with peak brightnesses greater than \mbox{$500$ mJy\: beam$^{-1}$}.

\section{Alternative Alignment Method}
\label{kevinappendix}

As an alternative approach to calibrating the image alignment we present a method currently under investigation based on cross correlation between epochs. In this section we present results from the \emph{R1} 850 $\mu$m reductions, however, this Cross Correlation method has proven to be successful for all four of the 850 $\mu$m reductions. 

The Cross Correlation method computes the cross correlation between a reference epoch to each succeeding science epoch:
\beq
cor(\text{R.A.},\text{Decl.}) = \sum_{\text{pixels}_x} \sum_{\text{pixels}_y} \mathcal{R}(x,y) \times \mathcal{S}(x-\text{R.A.},y-\text{Decl.})
\eeq 
where $\mathcal{R}$ is a reference epoch map, which we choose to be the first epoch from each region, $\mathcal{S}$ is a succeeding science epoch map to be aligned, and both maps have identical dimensions. 
The cross correlation of a reference epoch to a science epoch is the measure of how similar the two maps are as a function of the displacement of the science map relative to the reference map. If the two maps were identical and there were zero offset, then the measure of the cross correlation would be an auto correlation, where the peak $max\left( cor(\text{R.A.},\text{Decl.}) \right)$ resides at $(\text{R.A.},\text{Decl.}) = (0,0)$. The measured radial offset between the reference map and the science map is:
\begin{equation}
\text{Radial offset} = \sqrt{\left(\Delta \text{R.A.} \right)^2 + \left(\Delta \text{Decl.} \right)^2}
\end{equation}
where $\Delta \text{R.A.}$ and $\Delta \text{Decl.}$ are the angular offsets between $max\left( cor(\text{R.A.},\text{Decl.}) \right)$ and $(\text{R.A.},\text{Decl.}) = (0,0)$ in the right ascension and declination.

\begin{figure*}
	\centering
	\includegraphics[width=.8\textwidth]{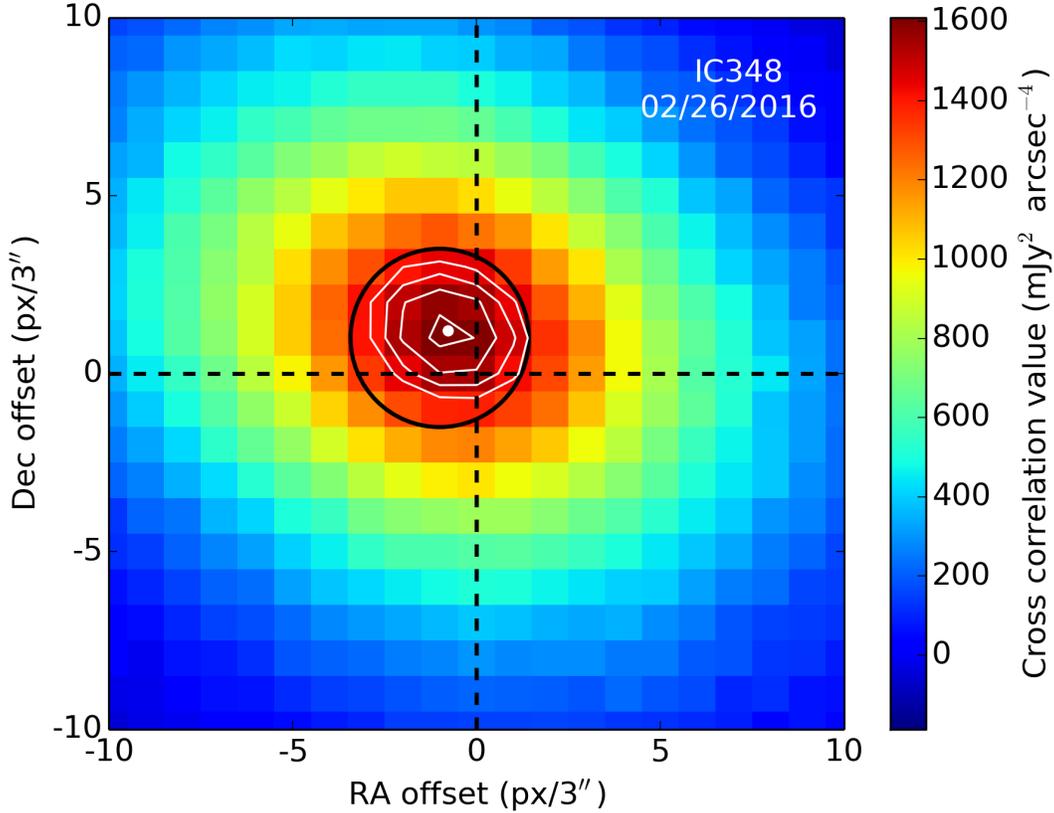} 
	\caption[Example cross correlation of IC348]{An example cross correlation of the IC348 region on 02/26/2016 compared to the reference image on 22/12/2015 for an \emph{R1} reduction at 850 $\mu$m. The black circle represents an effective $15^{\prime\prime}$ beam, the white contours represent the 2D Gaussian fit to the data, the white filled circle represents the position of the maximum of the fit, and the dashed black lines represent the zero offset position.}
	\label{fig:ex_cc}
\end{figure*}

To determine the position of $max\left( cor(\text{R.A.},\text{Decl.}) \right)$ a non-linear least squares regression is used to fit a 2D Gaussian to the inner $5\times5$ px$^2$ area, equivalent to a $15^{\prime\prime}$ beam at 850 $\mu$m \citep{dempsey2013} surrounding the most correlated pixel (e.g. Figure \ref{fig:ex_cc}). 
The uncertainty in the measured radial offset is estimated as the uncertainty of the 2D Gaussian fit.

The Cross Correlation method is advantageous to the {\sc{Gaussclumps}} method as the {\sc{Gaussclumps}} method considers a flux-limited sample, where it uses a list of bright compact small-scale structures, for which there could only be a few in some cases (e.g. IC348). Comparatively, the Cross Correlation method takes in consideration the entire map, including fainter and complex structures possibly missed by {\sc{Gaussclumps}}. 

\begin{figure*} 	
\centering
\subfloat{\label{}\includegraphics[width=9cm,height=7.8cm]{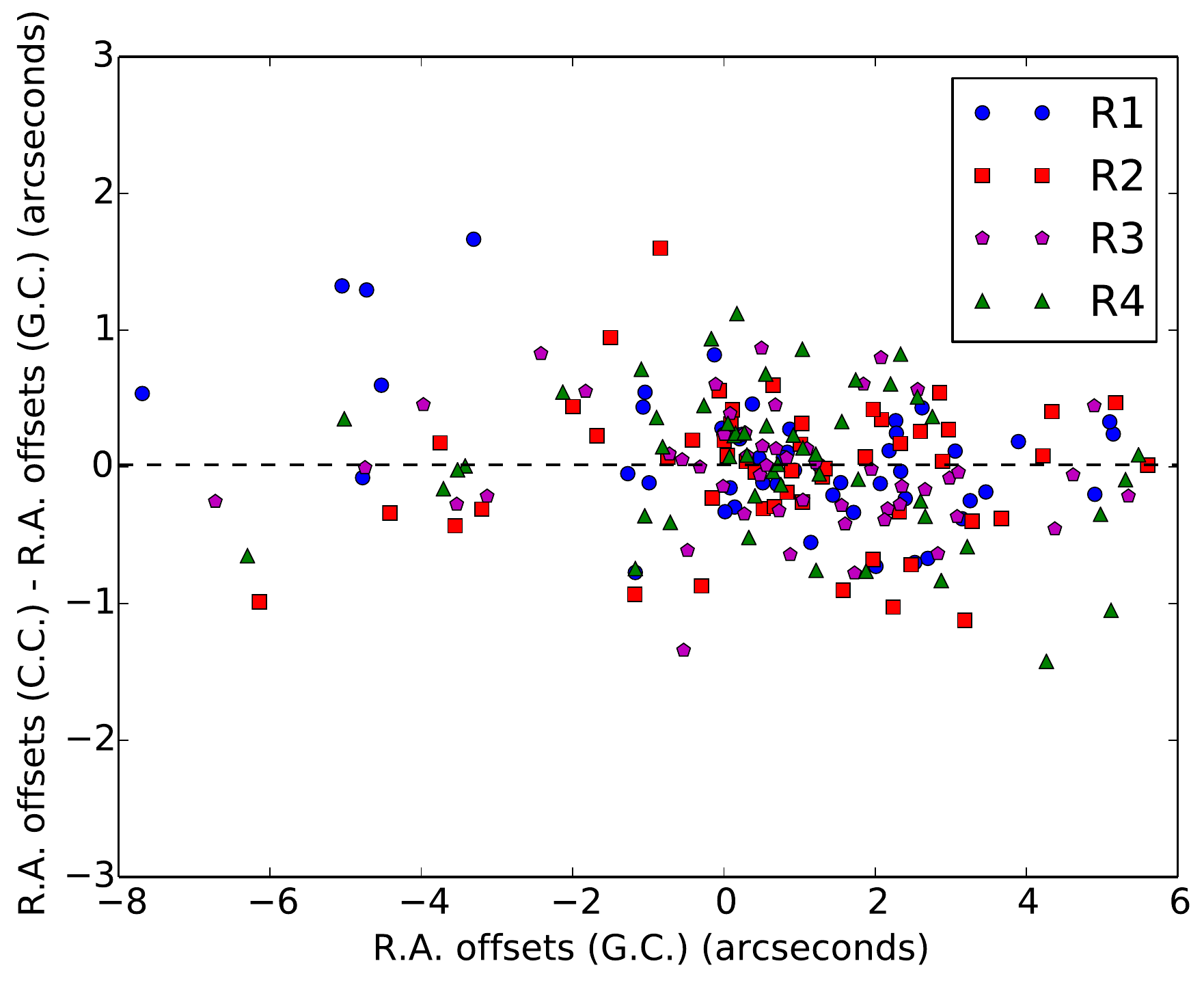}}
\subfloat{\label{}\includegraphics[width=9cm,height=7.8cm]{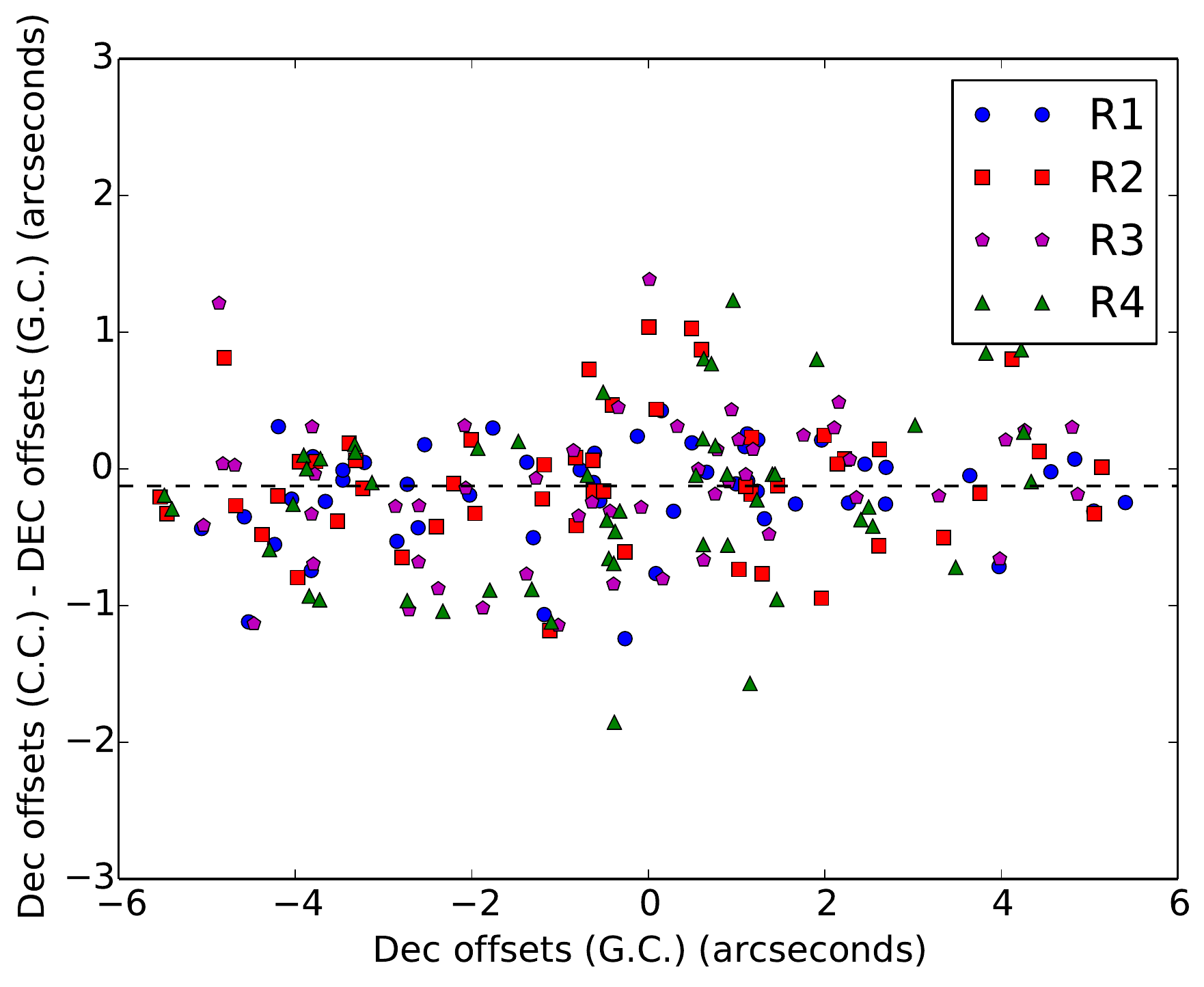}}
\caption{The right ascension (left) and declination (right) offsets derived using the Cross Correlation method compared with the offsets derived using the method described in the main paper. Compare with Figure \ref{align-app}.}
\label{fig:R1_K1_offset}
\end{figure*}

Positional offsets are measured for a total of 51 science epochs over all eight regions, and the median offset is $3.41\pm1.74\arcsec$, consistent with the initial position offsets measured from {\sc{Gaussclumps}} (see Figure \ref{fig:R1_K1_offset}). Subsequently, each science epoch is re-reduced with \emph{makemap} taking into account the derived offset relative to its reference map. Then, the same correlation and fitting method as described above is applied to the original data to the aligned maps in order to deduce any residual pointing uncertainty.

Using the {\sc{Gaussclumps}} method, we find comparable offset distributions for the unaligned maps. In Figure \ref{fig:R1_K1_offset}, we compare the right ascension and declination offsets derived using the Cross Correlation method (C.C.) with those derived using {\sc{Gaussclumps}} (G.C.) in a similar manner to Figure \ref{align-app} and find them to be consistent. The median residual offsets after alignment using the {\sc{Gaussclumps}} method is $\sim 0.6\arcsec$. 
Comparatively, the Cross Correlation method is able to self-consistently align maps to a scale $\sim20\times$ finer than the {\sc{Gaussclumps}} method with median residual offsets after alignment of $\sim0.03\arcsec$.
This alignment is $100\times$ smaller than a the pixel size at \mbox{850 $\mu$m}, and a factor of $\sim100\times$ better aligned than the telescope's pointing error. 

\begin{figure*}
	\centering
	\includegraphics[width=\textwidth]{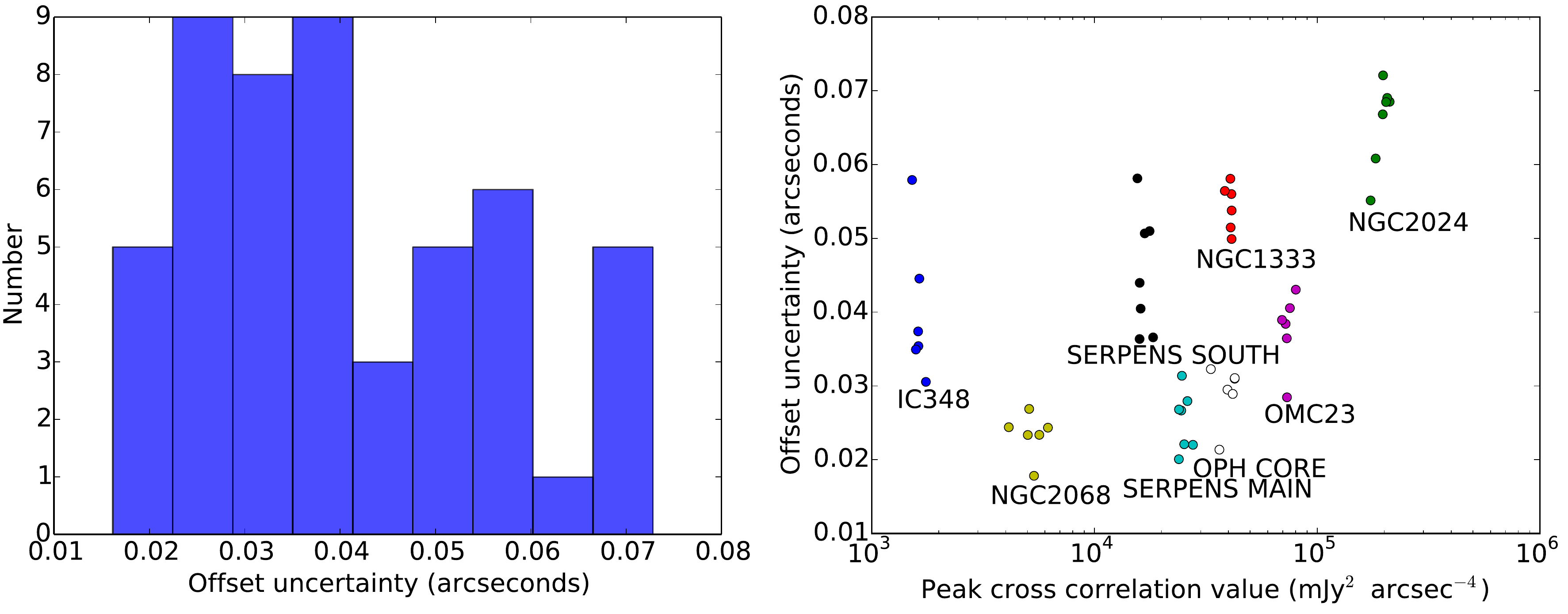}
	\caption[Properties of the Cross Correlation method]{\textit{Left:} Residual offset uncertainty distribution for aligned maps. \textit{Right:} Residual offset uncertainty as a function of maximum cross correlation value for aligned maps.}
	\label{fig:fitting_cc}
\end{figure*}

Although the median residual offset is $0.03\arcsec$, the accuracy of the alignment is limited by the uncertainty in the residual offset, which is typically larger than $\sim0.04\arcsec$ (see Figure \ref{fig:fitting_cc}). The uncertainty in the residual offset is limited by the uncertainty in the 2D Gaussian fit to the cross correlation, which is a result of the large spread of the cross correlation product (see Figure \ref{fig:ex_cc}). Therefore, this image alignment method is limited to a single iteration of the Cross Correlation method, as it will not improve on itself with succeeding iterations.

There do not seem to be any strong correlations between the measured residual offset and the maximum cross correlation value, nor any biases due to the fitting algorithm (see Figure \ref{fig:fitting_cc}). We find that the more bright compact small scale structure which resides within a region, the larger the peak cross correlation value.
While using {\sc{Gaussclumps}}, data in the NGC2024 region exhibits a higher uncertainty due to clustered sources mixing with larger-scale structure. Isolated, bright emission sources have less fitting uncertainties in {\sc{Gaussclumps}} (\citealt{stutzki1990}) and therefore produce the best alignments for the Transient Survey's current data. 
Thus, the {\sc{Gaussclumps}} method is biased towards having more accurate alignments for fields with compact bright sources embedded within small-scale structure, whereas the Cross Correlation method doesn't show strong correlation towards fields with either small-scale or irregular structures. As the survey matures, we will be exploring this alternate technique and refining our methodology to further improve our alignment calibration. 


\bibliography{varbib}

\end{document}